\documentclass[lettersize,journal]{IEEEtran}
\usepackage{amsmath,amsfonts}
\usepackage{algorithmic}
\usepackage{array}
\usepackage[caption=false,font=footnotesize,labelfont=rm,textfont=rm]{subfig}
\usepackage{textcomp}
\usepackage{stfloats}
\usepackage{url}
\usepackage{verbatim}
\usepackage{graphicx}
\usepackage{cite}
\usepackage{multirow}
\usepackage{makecell}
\usepackage{color}
\usepackage{amsmath,amssymb,amsfonts}
\makeatletter
\newcommand*\bigcdot{\mathpalette\bigcdot@{.5}}
\newcommand*\bigcdot@[2]{\mathbin{\vcenter{\hbox{\scalebox{#2}{$\m@th#1\bullet$}}}}}
\makeatother
\makeatletter

\newcommand{\Rmnum}[1]{\expandafter\@slowromancap\romannumeral #1@}
\makeatother
\usepackage[colorlinks,urlcolor=blue,linkcolor=blue,anchorcolor=blue,citecolor=blue]{hyperref}
\usepackage{bbm}
\usepackage[lined,boxed,commentsnumbered,ruled]{algorithm2e}
\usepackage[switch]{lineno}
\usepackage{subfig} 
\usepackage{subfloat}

\begin{document}
	%\title{Learning Color-aware Diffusion with Cross-Spectral Refinement for Underwater Image Restoration}
	\title{Color Correction Meets Cross-Spectral Refinement: A Distribution-Aware Diffusion for Underwater Image Restoration}
	%\title{Underwater Color Corrector: Distribution-Aware Diffusion Meets Cross-Spectral Refinement}
	%\title{Underwater Color Corrector: Distribution-Aware Diffusion Meets Cross-Spectral Refinement}
	%High-frequency
	\author{{Laibin~Chang, Yunke~Wang, Bo~Du,~\IEEEmembership{Senior Member,~IEEE}, Chang~Xu,~\IEEEmembership{Senior Member,~IEEE}}% <-this % stops a space
		\thanks{Laibin Chang and Bo Du are with the School of Computer Science, Institute of Artificial Intelligence, National Engineering Research Center for Multimedia Software, and Hubei Key Laboratory of Multimedia and Network Communication Engineering, Wuhan University, China. (E-mail: changlb666@whu.edu.cn; dubo@whu.edu.cn). \emph{(Corresponding author: Bo Du, Chang Xu.)}}
		\thanks{Yunke Wang and Chang Xu are with the School of Computer Science, The University of Sydney, Australia. (E-mail: yunke.wang@sydney.edu.au; c.xu@sydney.edu.au)}
	}
	
	\markboth{Journal of \LaTeX\ Class Files,~Vol.~14, No.~8, August~2021}%
	{Shell \MakeLowercase{\textit{et al.}}: A Sample Article Using IEEEtran.cls for IEEE Journals}
	
	\maketitle
	
	\begin{abstract}
		Underwater imaging often suffers from significant visual degradation, which limits its suitability for subsequent applications. While recent underwater image enhancement (UIE) methods rely on the current advances in deep neural network architecture designs, there is still considerable room for improvement in terms of cross-scene robustness and computational efficiency. Diffusion models have shown great success in image generation, prompting us to consider their application to UIE tasks. However, directly applying them to UIE tasks will pose two challenges, \textit{i.e.}, high computational budget and color unbalanced perturbations. To tackle these issues, we propose DiffColor, a distribution-aware diffusion and cross-spectral refinement model for efficient UIE. Instead of diffusing in the raw pixel space, we transfer the image into the wavelet domain to obtain such low-frequency and high-frequency spectra, it inherently reduces the image spatial dimensions by half after each transformation. Unlike single-noise image restoration tasks, underwater imaging exhibits unbalanced channel distributions due to the selective absorption of light by water. To address this, we design the Global Color Correction (GCC) module to handle the diverse color shifts, thereby avoiding potential global degradation disturbances during the denoising process. For the sacrificed image details caused by underwater scattering, we further present the Cross-Spectral Detail Refinement (CSDR) to enhance the high-frequency details, which are integrated with the low-frequency signal as input conditions for guiding the diffusion. This way not only ensures the high-fidelity of sampled content but also compensates for the sacrificed details. Comprehensive experiments demonstrate the superior performance of DiffColor over state-of-the-art methods in both quantitative and qualitative evaluations. %The code is available at \href{https://github.com/LaibinChang/UFDM.git}{https://github.com/LaibinChang/UFDM}.
	\end{abstract}
	
	
	\IEEEpeerreviewmaketitle
	
	\section{Introduction}\label{Introduction}
	
	%\begin{figure}[!htp]
		%\setlength{\abovecaptionskip}{0.0cm}
		%\setlength{\belowcaptionskip}{-0.2cm}
		%\centering
		%\includegraphics[width=0.48\textwidth]{Figs/1-0The_First_Picture.pdf}
		%\caption
		%{Qualitative and quantitative comparisons with state-of-the-art (SOTA) methods on the underwater image benchmark (UIEB). (a) Our method achieves the most homogeneous visual perception and does not introduce blue-green artifacts around the fish, as seen in other methods (marked with red arrows). (b) Our method is superior to SOTA methods in terms of PSNR, SSIM, and Inference Time. More details of the evaluation can be found in the experimental section.}
		%\label{1-0The_First_Picture}
	%\end{figure}
	
\begin{figure}[!tp]
		\setlength{\abovecaptionskip}{-0.1cm}
		\setlength{\belowcaptionskip}{-0.3cm}
		\centering
		\includegraphics[width=0.5\textwidth]{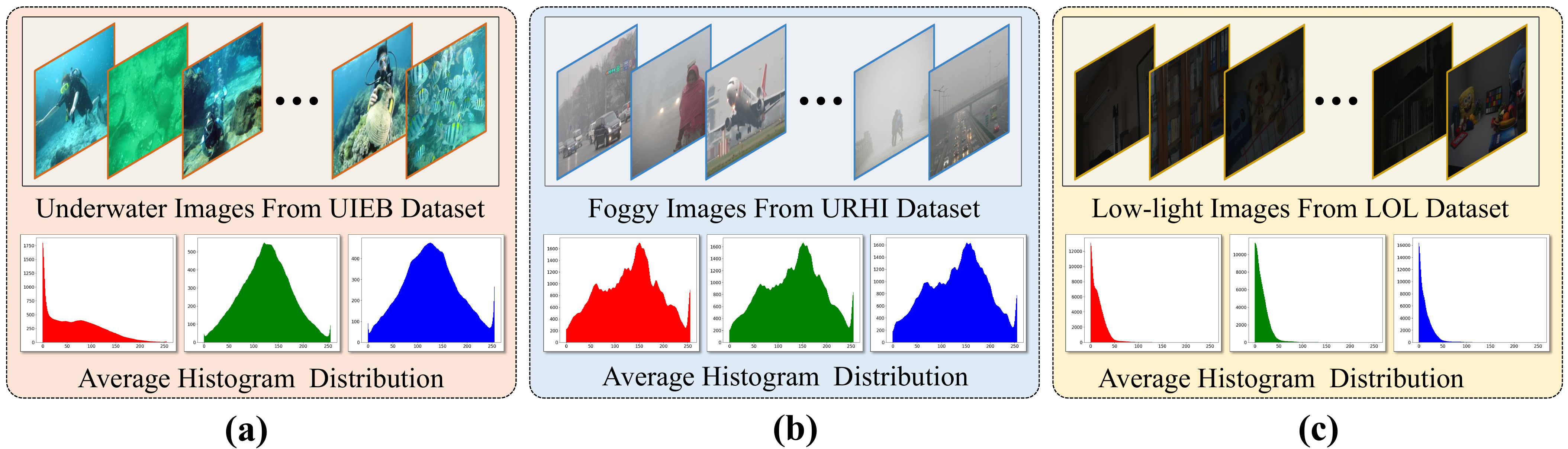}
		\caption
		{Average histogram distribution analysis of R-G-B three channels on underwater images, foggy images, and low-light images. (a) the top shows some examples from the underwater benchmark dataset UIEB \cite{Li2020UIEB}, while the bottom is the average histogram distribution of all images in the UIEB dataset. (b) displays examples and average histogram distribution related to the foggy dataset URHI \cite{Li2018URHI}, while (c) illustrates examples and histogram distribution of the low-light dataset LOL \cite{wei2018deep}. Compared with foggy and low-light scenes, there are significant disparities in the average channel distribution of underwater scenes.}
		\label{1-1Histogram_Distribution_Comparison}
	\end{figure}

	\IEEEPARstart{U}{nderwater} clear imaging plays an irreplaceable role in coral reef segmentation \cite{Zheng2024Coralscop}, marine object detection \cite{Zhang2024Fantastic, Liu2024Detection}, and autonomous underwater vehicles \cite{Cai2023AUV, Wang2024Uierl}. However, raw images captured directly from underwater cameras often suffer from various types of degradation, including color deviation, low contrast, and blurred details, caused by the selective absorption and strong scattering of light in water \cite{Jiang2021Underwater, Lin2024Underwater, Guo2022Underwater, Liu2024Underwater, Li2024Learning}. Degraded images with these imperfections are not only visually unappealing, but may also impede auxiliary operations for underwater exploration tasks that rely on vision to perceive the surrounding environment. As shown in Fig. \ref{1-1Histogram_Distribution_Comparison}, compared with single-noise image restoration tasks (low-light enhancement \cite{Wu2023LLE, Li2024LLE, Ma2023LLE, Jiang2022LLE} and defogging \cite{Patil2023Defoging, Sun2024Defoging}), underwater image enhancement (UIE) necessitates removing various noise interferences and equalizing the information distribution across three color channels.

\begin{figure}[!tp]
		\setlength{\abovecaptionskip}{-0.1cm}
		\setlength{\belowcaptionskip}{-0.3cm}
		\centering
		\includegraphics[width=0.48\textwidth]{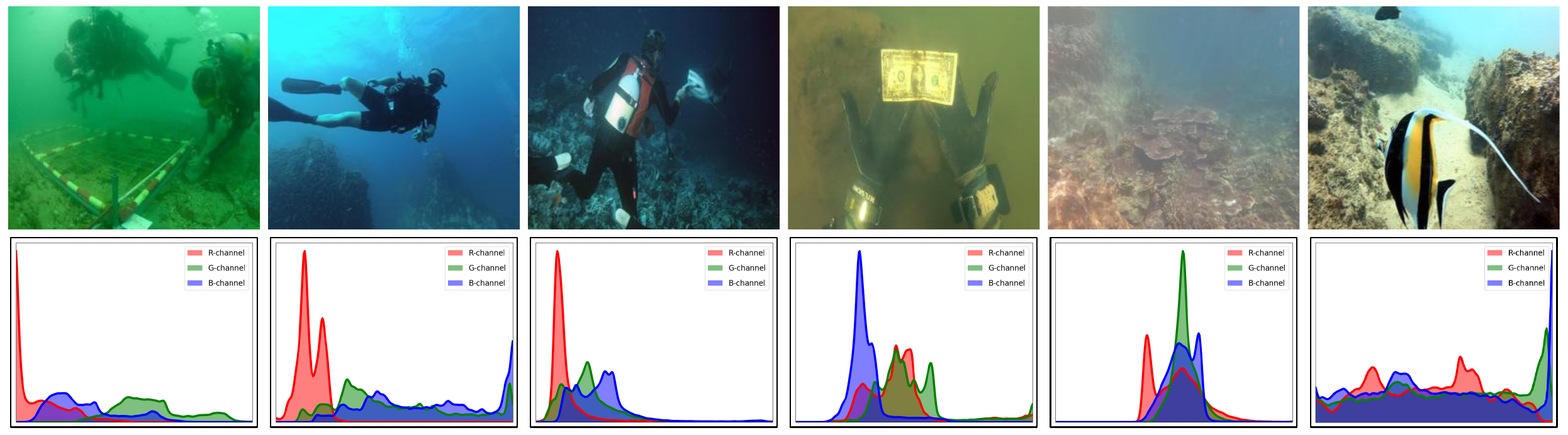}
		\caption
		{Underwater images with different degradation types have various information distributions of R-G-B three-color channels.}
		\label{1-2Various_Degraded_Types}
		\vspace{-0.3cm}
	\end{figure}

	Recently, numerous UIE methods derived from different inspirations have been proposed, ranging from physics-based \cite{Zhang2022MLLE, Wang2023UCD, Song2022ERH, Chang2023ISPRS, Zhou2024Pixel, Hou2023Non-uniform} to deep learning-based \cite{Jiang2022Two_step, Zhou2024HCLR, Xue2023IntrinsicDF, Huang2023Semi-UIR, Qi2022JointLearning, Qi2022Sguie, Liu2023WSDS_GAN, Qing2024Unformer, Qing2024Diffuie, Du2025Uiedp}.
	%several state-of-the-art UIE methods are evaluated in Fig. \ref{1-0The_First_Picture}.
	These physics-based methods estimate the desired enhancement result from a single degraded image based on underwater imaging priors or pixel-value modification techniques. Despite their success, the effectiveness of these methods depends heavily on the accuracy of hand-crafted priors to improve the visibility of underwater images. However, these priors with invariant parameters are not sufficient to enhance the severely degraded images when confronted with variable underwater environments \cite{Huang2023Semi-UIR, Jiang2024Towards}. Instead of designing hand-crafted priors to improve the visibility of underwater images, deep learning-based solutions can provide relatively satisfactory enhancement results due to their superior data-driven capabilities \cite{Jiang2024Perception, Qing2024Diffuie, Liu2024CCL_Net}. However, there are several non-negligible problems with existing deep learning-based methods: 1) The GAN-based UIE methods with adversarial loss occasionally introduce artifacts not present in the clean reference image, leading to noticeable distortion. 2) These methods usually learn non-linear restoration mappings in the raw pixel space of underwater images, with limited exploration and utilization of spatial properties in the frequency domain of the image, resulting in an ineffective generation of high-quality images. 3) Most methods overlook the distribution disparities across the R-G-B channels in underwater images (as described in Fig. \ref{1-2Various_Degraded_Types}) when designing their model architecture.
	
	%Consequently, these intra-domain and inter-domain gaps of different degradation types across underwater scenes pose significant challenges for deep learning-based models in learning non-linear restoration mappings.
	
	Diffusion models (DMs) have recently garnered increasing attention due to their impressive performance in image restoration, especially in terms of noise robustness. However, when directly applying DMs to underwater image restoration tasks, they usually require long-time inference to achieve a high-quality mapping from random Gaussian noise to the target image, especially for large-size underwater images. Recently, inspired by the Denoising Diffusion Implicit Models (DDIM) \cite{Song2020DDIM} to reduce sampling steps, some DM-based methods \cite{Tang2023DM-based, Lu2023SpeedDM} have incorporated the skip-sampling strategy to reduce the iterative steps in the denoising process of underwater images. Unfortunately, this manner not only irreversibly sacrifices the high-fidelity of its inherent property to a certain extent, but also fails to exploit spectral information for image restoration. There is a wavelet-based diffusion solution in \cite{Zhao2024Wavelet} that demonstrates the effectiveness of decreasing inference time by halving the underwater image resolution after each transformation. However, the proposed model employs a two-stage scheme that establishes two separate diffusion models to restore low-frequency and high-frequency components, which inadvertently increases both the model's parameter count and computational overhead. In summary, while previous conditional DMs suffice for single-noise restoration tasks in regular images, it should be noted that underwater environments necessitate heightened attention in restoring the global degradation disturbances induced by light attenuation during the reverse denoising process, especially for the color distribution correction and sacrificed detail refinement.
	
	%it should be noted that the available underwater image datasets are not yet able to encompass the various degradation types of underwater imaging. Thus, underwater environments necessitate heightened attention in restoring the global degradation disturbances induced by light attenuation during the reverse denoising process, especially for the color distribution correction and sacrificed detail refinement.
	%Unfortunately, the degradation types of underwater images are more variable, and underwater scenes require more attention to recovering the global degradation perturbations caused by light attenuation during the inverse denoising process, especially for color correction and detail refinement.
	\begin{figure*}[!htp]
		\setlength{\abovecaptionskip}{0.0cm}
		\setlength{\belowcaptionskip}{-0.2cm}
		\centering
		\includegraphics[width=0.96\textwidth]{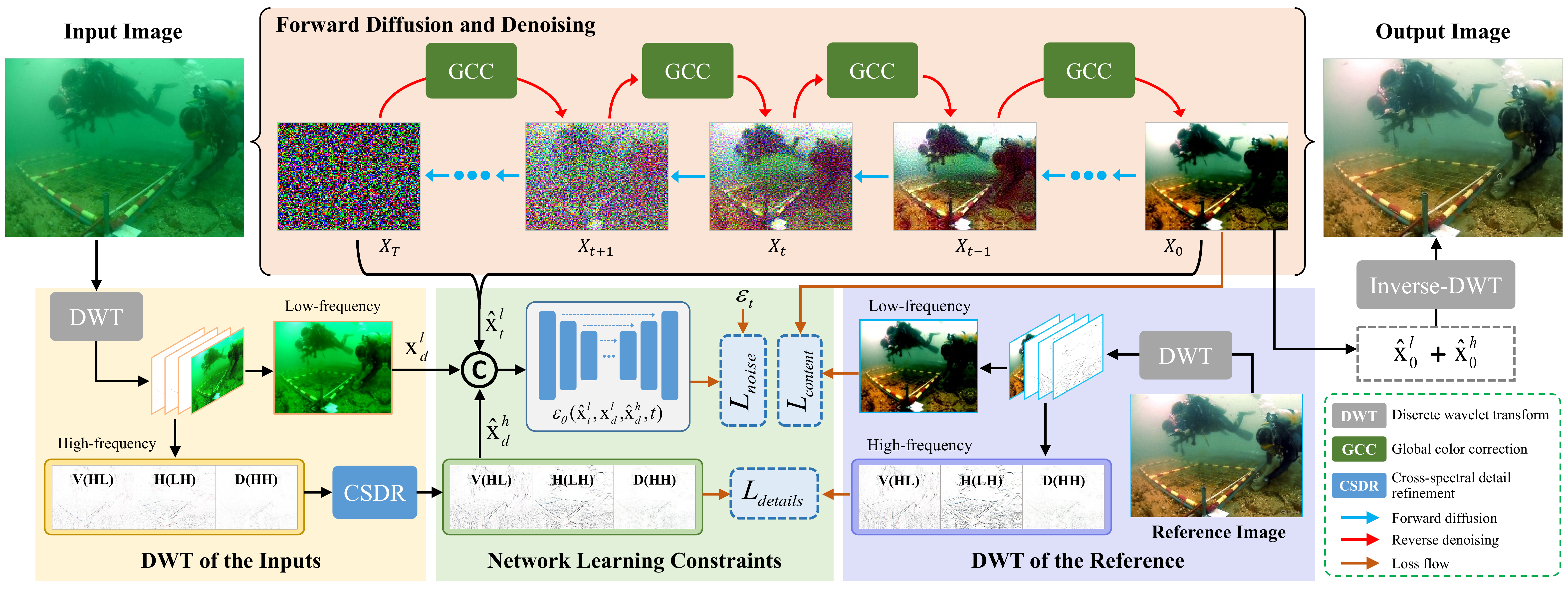}
		\caption
		{Overview of the proposed DiffColor. Instead of the raw pixel space, the degraded and reference images are converted into the wavelet domain, yielding one low-frequency component that characterizes the principal content of the image and three high-frequency spectra that represent the image details. For the various color shifts caused by underwater selective absorption, the global color correction (GCC) module is employed to modulate the color information distribution of low-frequency component $\mathrm{\hat{x}}^{l}_{t}$ during the denoising. For the sacrificed image details caused by underwater medium scattering, the cross-spectral detail refinement (CSDR) module is adopted to enhance the high-frequency details $\mathrm{\hat{x}}^{h}_{d}$, which are then integrated with the low-frequency component $\mathrm{x}^{l}_{d}$ as input conditions for the noise estimation network.  Finally, the enhanced image is obtained by inverting the generated low-frequency component and three refined high-frequency spectra.}
		\label{1-1The_Overall_Framework}
		% \vspace{-0.3cm}
	\end{figure*}

	Motivated by the aforementioned analysis, we propose an underwater color corrector (DiffColor) that incorporates distribution-aware diffusion and cross-spectral refinement for efficient underwater image restoration. As depicted in Fig. \ref{1-1The_Overall_Framework}, we initially employ the discrete wavelet transform to transform the degraded image into the wavelet domain, yielding one low-frequency component that characterizes the principal content of the image, along with three high-frequency spectra carrying local details. Instead of operating in the raw image domain with a skip sampling strategy, diffusing on the low-frequency component can reduce the inference time by halving the spatial dimensions after each transformation. This manner also effectively avoids error information accumulation that may occur with non-uniform sampling. In response to the various color shifts caused by the selective absorption of light in water, we design the global color correction (GCC) module to balance the information distribution across R-G-B three channels in the low-frequency component, thereby preventing the global degradation disturbances that commonly arise during the denoising process. For the sacrificed image details caused by underwater medium scattering, we further propose the cross-spectral detail refinement (CSDR) module to enhance the high-frequency details, which are integrated with the low-frequency signal as input conditions. This way not only maximizes the generative capability of the diffusion model but also compensates for the details sacrificed due to underwater scattering.%It differs from the previous DM-based UIE methods, which relied solely on degraded images as conditions.
	
	  %color shifts during the reverse denoising process.
	 
	%Compared to natural images captured in landscapes, the two main interfering factors in underwater imaging are the scattering of light by the water medium and its selective absorption of light.
	
	%and then combined with the low-frequency signal as input conditions for the noise estimation network, ensuring the high-fidelity of the restored content even if the reverse starts from a randomly sampled noise.
	%This way not only maximizes the generative capability of the diffusion model but also compensates for the details sacrificed due to underwater scattering. 
	%which combines the impressive generative advantages of the diffusion model with the dimensionality reduction of the discrete wavelet transform, achieving fast inference speed in the wavelet domain.
	%Our DiffColor model employs a wavelet-guided diffusion scheme, independently achieving color correction and detail refinement in both low-frequency and high-frequency spectra. 
	
	Our key contributions are summarized as follows:
	\begin{itemize}
		\item  We propose distribution-aware DiffColor to modulate the denoising and avoid global degradation, which is committed to addressing the unbalanced channel distribution caused by underwater light absorption.
		%Considering the various chromatic aberrations originated from underwater light absorption, which manifests as unbalanced histogram distribution across the R-G-B channels. We propose the GCC module to modulate the reverse denoising in the diffusion model, thus preventing global degradation caused by color shifts.
		
		%Unlike the restoration tasks for single-noise images (dehazing and low-light enhancement), 
		
		\item We design the CSDR module to refine image details by exploiting the cross-directional complementarity between different high-frequency spectra and treating them as conditional signals to guide the diffusion.
		%Instead of previous DM-based UIE solutions that only conditioned with low-quality images, w
		
		\item Comprehensive experiments demonstrate that DiffColor outperforms existing state-of-the-art UIE solutions in both qualitative and quantitative evaluations.
		
	\end{itemize}
	
	\section{Related Work}
	\subsection{Underwater Image Enhancement}
	Underwater image enhancement has garnered widespread attention and can be broadly categorized into physics-based and deep learning-based methods. Early physics-based methods \cite{Akkaynak2018Revised, Li2020UnderwaterPrior, Zhou2023Underwater} focused on designing hand-crafted priors to estimate unknown quantities in underwater imaging models, including background light and transmission parameters. Given the similarities between underwater degradation and foggy scenes, several revised DCP-based (Dark Channel Prior) methods have been proposed for underwater image restoration, such as UDCP \cite{Drews2013UDCP}, GDCP \cite{Peng2018GDCP}, and GUDCP \cite{Liang2022GUDCP}. However, there are many disturbing factors affecting underwater imaging, and it is difficult to accurately estimate these environmental parameters. The remaining methods improved the visual perception by directly manipulating image pixel values with well-designed techniques, regardless of the underwater degradation mechanism, including fusion-based \cite{Ancuti2018Fusion-based, Song2022ERH, Chang2023ISPRS, Zhang2022ACDC}, Retinex-based \cite{Fu2014Retinex, Zhuang2022Retinex, Zhang2022Retinex}, and histogram equalization-based \cite{Garg2018CLAHE, Huang2018RGHS, Zhou2023MultiSubhistogram}, \textit{etc}. Despite their success, these methods tend to adopt invariant empirical parameters whose poor generalisability cannot cope with complex and changeable underwater scenes.
	
	Deep learning-based methods concentrated on enhancing degraded images by autonomously learning non-linear restoration mappings from paired underwater image datasets. Specifically, many generative adversarial networks (GANs)-based methods have been proposed for underwater image enhancement tasks, such as UGAN \cite{Fabbri2018UGAN}, WaterGAN \cite{Li2018WaterGAN}, FUnIE-GAN \cite{Islam2020EUVP}, UWGAN \cite{Guo2020MultiscaleGAN}, and TwinGAN \cite{Liu2022TwinGAN}. However, GAN-based restoration methods with adversarial losses often introduce artifacts that are not present in the pre-processed images during unstable training, and may even fail due to mode collapse. Furthermore, Wang \emph{et al.} \cite{Sun2022RL, Wang2023TargetPerception} proposed the reinforcement learning methods based on Markov decision process (MDP) for underwater image enhancement. Inspired by the revised underwater imaging models, several deep learning methods \cite{Wang2017Imaging_model, Chen2020Perceptualmodel, Song2023Dual_model} have been proposed for reconstructing clear underwater images by estimating the ambient-light and direct-transmission parameters, which are superior to the prior estimation based on a single image. Faced with the noticeable gap between the degradation distributions of real-world underwater images, some end-to-end domain adversarial learning networks have been proposed to promote their adaptive enhancement capabilities, such as TUDA \cite{Wang2023TUDA}, Two-SDA \cite{Jiang2022Two_step}, UICoE-Net \cite{Qi2022JointLearning}, \textit{etc}. Although these methods attain domain adaption to some extent, most of them have low execution efficiency due to the two-stage model, leaving much room for improvement to meet the underwater real-time processing requirements.

	\subsection{Diffusion Model-based Image Restoration}
	Conditional diffusion-based models have yielded excellent performance with their powerful generative capabilities and have been applied in a variety of image restoration tasks, including low-light enhancement \cite{Yin2023CLEDiff, Ooi2023LLDE}, super-resolution \cite{Gao2023SRDM}, inpainting \cite{Lugmayr2022RepaintDM}, deblurring \cite{Wu2023DebluringDM}, \textit{etc}. In the field of underwater image restoration, Lu \emph{et al.} \cite{Lu2023UW-DDPM} proposed an underwater image enhancement method called UW-DDPM, which utilizes two U-Net networks to perform denoising and distribution transforms. Although UW-DDPM is superior to the GAN-based UIE methods, the multiple forward and reverse passes throughout the whole network usually require a substantial amount of computational resources. Inspired by the idea of reducing sampling in \cite{Song2020DDIM}, Tang \emph{et al.} \cite{Tang2023DM-based} introduced a lightweight transformer-based denoising diffusion network for underwater image enhancement and adopted a skip sampling strategy to reduce backward iteration steps. Then, Lu \emph{et al.} \cite{Lu2023SpeedDM} accelerated the inference process of UIE by modifying the initial sampling distribution and reducing the number of iterations in the denoising phase. However, these diffusion-based underwater image enhancement methods only operate in the raw image domain, and have limited exploration and utilization of the spatial characteristics in the image frequency domain. Recently, Zhao \emph{et al.} \cite{Zhao2024Wavelet} proposed a wavelet-based diffusion model named WF-Diff to enhance underwater images, which includes the wavelet-based Fourier information interaction network and frequency residual diffusion adjustment network. However, WF-Diff is a two-stage scheme that initially utilizes frequency information exchange to obtain coarse results, and then designs two separate diffusion models to respectively restore low-frequency and high-frequency components, inevitably exacerbating the model's parameters and computational burden.
	
	%There is an imminent requirement for an efficient UIE model to handle real-time underwater video processing in practical scenarios.
	
	%In addition, there are many interfering factors that affect underwater imaging, making it challenging to accurately calculate the parameters of these complex environments using a single diffusion model.
	
	%To address these limitations, this paper introduces an efficient underwater image enhancement method based on the wavelet-guided diffusion model. This method not only relies on the dimensionality reduction advantages of wavelet transform to improve the inference speed of the model, but also incorporates high-frequency refinement and color correction modules to mitigate global degradation caused by inherent underwater properties such as selective absorption and medium scattering.
	
	\section{Methodology}
	\subsection{Motivation Stems From Two Challenges}
	\textbf{Heavy Computational Burdens and Frequency Domain Exploitation.}
	\begin{figure}[!ht]
		\setlength{\abovecaptionskip}{0.0cm}
		\setlength{\belowcaptionskip}{-0.2cm}
		\centering
		\includegraphics[width=0.486\textwidth]{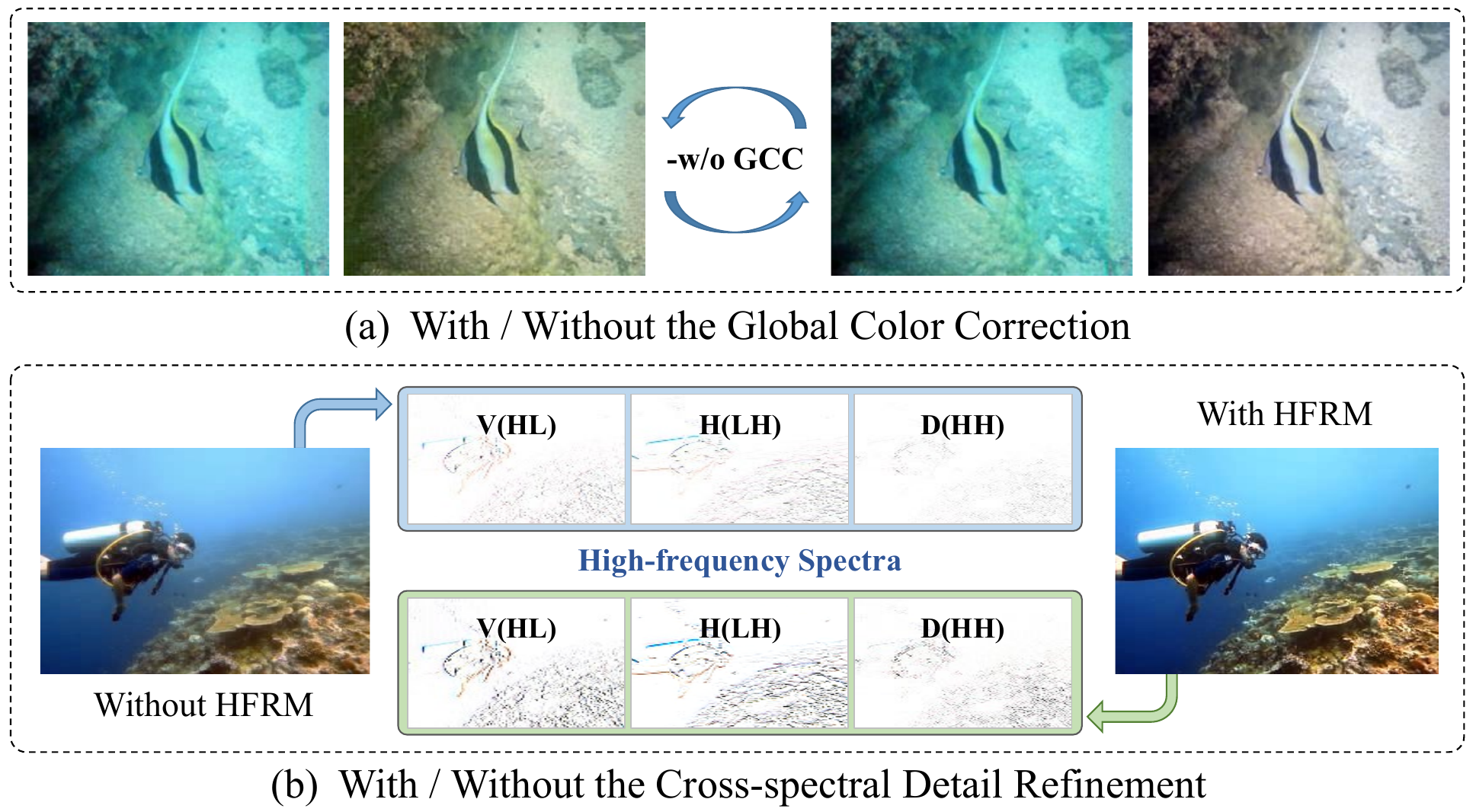}
		\caption
		{Ablation comparison regarding the necessity of the GCC and CSDR modules. (a) The GCC module corrects the global degradation caused by underwater selective absorption. (b) The CSDR module refines the image details sacrificed due to underwater medium scattering.}
		\label{2-1Without_GCC_or_CSDR}
		\vspace{-0.2cm}
	\end{figure}
	%Despite the commendable achievements of diffusion models in image restoration, they often demand substantial computational resources, especially for high-resolution panoramas. 
	Reducing the resolution of large-scale images with a down-sampling operation can significantly decrease the computational resources required by the diffusion model. However, this down-sampling way will irreversibly lead to the loss of image information during the inverse process. The Discrete Wavelet Transform (DWT) can effectively halve the resolution of the input image after each transformation, converting it into a low-frequency component and three high-frequency details without sacrificing information. In addition, learning non-linear restoration mappings in the raw pixel space of underwater images is limited, while exploring and utilizing spatial characteristics in the frequency domain can aid in generating high-quality images. Hence, we utilize the DWT to transform the degraded image into the wavelet domain, yielding a halved low-frequency component and three complementary high-frequency spectra.
	
	%\subsubsection{Details Degradation Caused by Medium Scattering.}
	\textbf{Global Degradation Caused by Underwater Absorption and Scattering.}
	In contrast to natural images captured from landscapes, underwater image degradation is mainly interfered with selective absorption and medium scattering in water. Specifically, underwater selective absorption leads to color deviation in degraded images, while medium scattering causes their details to appear blurred. Taking into account both color correction and detail refinement in a limited number of sampling steps is a significant challenge for diffusion models, especially for dealing with underwater images with more severe degradation. As shown in Fig. \ref{2-1Without_GCC_or_CSDR}, we not only need to correct for biased colors with the Global Color Correction (GCC) module, but also enrich the sacrificed details with the Cross-Spectral Detail Refinement (CSDR) in the whole restoration process.
	
	%these color shifts cannot be corrected using diffusion models alone \cite{Zhou2023PyramidDM}.
	\subsection{Frequency-guided Conditional Diffusion}
	Considering the generative capability of the diffusion model and the dimensionality reduction advantages of the discrete wavelet transform, we incorporate these two designs and propose the underwater distribution-aware diffusion model (DiffColor) to enhance underwater images. As depicted in Fig. \ref{1-1The_Overall_Framework}, we perform the diffusion operations in the wavelet domain instead of the image space by transforming the degraded image into low-frequency and high-frequency components. In addition, we adopt the GCC module to correct the unbalanced color distribution caused by selective absorption, and introduce the CSDR module to refine the local texture details sacrificed due to medium scattering.
	
	Given a degraded underwater image $X_{d}\in \mathbb{R}^{H\times W\times 3}$ and its corresponding reference $X_{0}\in \mathbb{R}^{H\times W\times 3}$, we utilize the two-dimensional DWT with Haar wavelet to transform them into wavelet spectra with four sub-bands, respectively. The Haar wavelet consists of a low-pass filter $L$ and a high-pass filter $H$, and their expressions are as follows:
	\begin{equation}
	L=\frac{1}{\sqrt{2}}\begin{bmatrix} 1& 1\end{bmatrix},\ \ H=\frac{1}{\sqrt{2}}\begin{bmatrix}
	-1& 1\end{bmatrix}.
	\end{equation}
	
	The two filters $L$ and $H$ are used for constructing four kernels with stride 2, including $LL^{T}$, $HL^{T}$, $LH^{T}$, and $HH^{T}$. These kernel functions transform the input images $X_{d}$ and $X_{0}$ into a low-frequency component and three high-frequency spectra, which are expressed as follows:
	\begin{equation}
	\{\mathrm{x}^{l}_{d},\mathrm{x}^{h}_{d},\mathrm{x}^{l}_{0},\mathrm{x}^{h}_{0}\} = DWT(X_{d},X_{0}),
	\end{equation}
	where $\{\mathrm{x}^{l}_{d},\mathrm{x}^{l}_{0}\}\in \mathbb{R}^{\frac{H}{2}\times \frac{W}{2}\times 3}$ represent the low-frequency components, and $\{\mathrm{x}^{h}_{d},\mathrm{x}^{h}_{0}\}\in \mathbb{R}^{\frac{H}{2}\times\frac{W}{2}\times 9}$ represent the high-frequency spectra, encompassing three directions of the vertical $\{V_d,V_0\}\in \mathbb{R}^{\frac{H}{2}\times\frac{W}{2}\times 3}$, horizontal $\{H_d,H_0\}\in \mathbb{R}^{\frac{H}{2}\times\frac{W}{2}\times 3}$, and diagonal $\{D_d,D_0\}\in \mathbb{R}^{\frac{H}{2}\times\frac{W}{2}\times 3}$, respectively. The low-frequency component $\mathrm{x}^{l}_{d}$ is involved in the inference process of the diffusion model to restore the global content, while the high-frequency spectral bands $\mathrm{x}^{h}_{d}$ serve as indispensable conditions to enrich the local details.
	
	%The low-frequency component $\mathrm{x}^{l}$ of an image can be expressed as follows:
	%\begin{equation}
	%\mathrm{x}^{l}(x,y) =\frac{\sum_{i=0}^{M-1}\sum_{j=0}^{N-1}I(i,j)\varphi_{x,y}(i,j)}{\sqrt{M\times N}},
	%\end{equation}
	%where $M$ and $N$ denote the width and height of the input image $I(i,j)$, respectively. $\varphi_{x,y}(i,j)$ is a 2-dimensional Gaussian function with the constraint of $(x,y)=2^{n}-1$. In addition, the high-frequency spectra are calculated as follows:
	
	%\begin{equation}
	%\mathrm{x}^{h}(x,y) =\frac{\sum_{i=0}^{M-1}\sum_{j=0}^{N-1}I(i,j)\phi_{x,y}(i,j)}{\sqrt{M\times N}},
	%\end{equation}
	%where $\phi_{x,y}(i,j)$ represents the 2-dimensional wavelet function, and $\mathrm{x}^{h}\in \left\{ V, H, D \right\}$ represents the vertical, horizontal, and diagonal high-frequency components, respectively.
	
	During the forward diffusion, we gradually add Gaussian random noise to the low-frequency component $\mathrm{x}^{l}_{d}$ in $T$ steps, following $q(\mathrm{\hat{x}}^{l}_{t}|\mathrm{x}^{l}_{0})=\mathcal N\left(\mathrm{\hat{x}}^{l}_{t};\sqrt{\bar{\alpha}_{t}}\mathrm{x}^{l}_{0},(1-\bar{\alpha}_{t})\mathrm{\textit{\textbf{I}}}\right), t=1,2,\cdots, T$. Considering that the spatial dimension of the low-frequency component is halved after each transformation, diffusion on this component effectively reduces the computational burden of the noise estimation network. This way also avoids the error-prone information accumulation associated with a multi-scale sampling strategy. Besides, we adopt a lightweight cross-spectral detail refinement (CSDR) to enhance the high-frequency spectral distribution between the degraded image and the clearer version, with the following loss function expressed as follows:
	
	\begin{equation}\label{Details_loss_function}
	\mathcal{L}_{details}=\sum_{k=1}^{K}\left\|\{\hat{V}^{k}_d,\hat{H}^{k}_d,\hat{D}^{k}_d\} - \{V^{k}_0,H^{k}_0,D^{k}_0\}\right\|^2,
	\end{equation}
	where $K$ represents the total number of wavelet transforms and $k$ represent $k$-th wavelet scale.
	
	Instead of the traditional diffusion model only taking the degraded image X as a condition, we jointly incorporate the sampled results $\mathrm{\hat{x}}^{l}_{t}$, the low-frequency component $\mathrm{x}^{l}_{d}$, and the enriched high-frequency coefficients $\mathrm{\hat{x}}^{h}_{d}$ as conditional inputs for the denoising network $\epsilon_{\theta}(\mathrm{\hat{x}}^{l}_{t},\mathrm{x}^{l}_{d},\mathrm{\hat{x}}^{h}_{d},t)$, ensuring the high-fidelity of the restored content even if starting from a randomly sampled noise. This way not only maximizes the generative capability of the diffusion model but also compensates for the details sacrificed due to underwater scattering. Based on this, the objective loss function is defined as:
	\begin{equation}\label{Revised_noise_loss_function}
	\mathcal{L}_{noise}=\mathbbm{E}_{\mathrm{x}_{0},\epsilon_{t}\sim\mathcal{N}(0,\mathrm{\textit{\textbf{I}}})}\|\epsilon_{t}-\epsilon_{\theta}(\mathrm{\hat{x}}^{l}_{t},\mathrm{x}^{l}_{d},\mathrm{\hat{x}}^{h}_{d},t)\|^{2},
	\end{equation}
	where $t=1,2,\cdots, T$. $\epsilon_{t}$ and $\epsilon_{\theta}$ represent the noise added at time $t$ and the predicted noise, respectively.
	
	\begin{algorithm}[!ht]
		\caption{The Training of DiffColor}
		\label{Training}
		\textbf{Input}: Degraded and clear images, denoted as $\mathrm{X}_{d}$ and $\mathrm{X}_{0}$; Noise schedule $\beta_{t}$; Noise estimation network $\theta$; Diffusion steps $T$ and sampling steps $S$.\\
		\textbf{Repeat}
		\begin{algorithmic}[1] %[1] enables line numbers
			\STATE \textit{\# Discrete wavelet transform}
			\STATE $\{\mathrm{x}^{l}_{d},\mathrm{x}^{h}_{d},\mathrm{x}^{l}_{0},\mathrm{x}^{h}_{0}\}=DWT(\mathrm{X}_{d},\mathrm{X}_{0})$
			\STATE \textit{\# Refine high-frequency components}
			\STATE
			$\mathrm{\hat{x}}^{h}_{d}=CSDR(\mathrm{{x}}^{h}_{d})$
			\STATE \textit{\# Forward diffusion process}
			\STATE Sample $t\sim$ Uniform$\{1,2,\cdots, T\}$
			\STATE Sample $\epsilon_{t}\sim \mathcal{N}(0,\mathrm{\textit{\textbf{I}}})$
			\STATE $\mathrm{\hat{x}}^{l}_{t}=\sqrt{1-\beta_{t}}\mathrm{x}^{l}_{t-1}+\beta_{t}\epsilon_{t}$
			\STATE Perform a single gradient descent step on\\
			$\nabla_{\theta}\|\epsilon_{t}-\epsilon_{\theta}(\mathrm{\hat{x}}^{l}_{t},\mathrm{x}^{l}_{d},\mathrm{\hat{x}}^{h}_{d},t)\|^{2}$
			\STATE \textit{\# Reverse denoising process}
			\STATE Sample ${\mathrm{x}}_{T}\sim \mathcal N(0,\mathrm{\textit{\textbf{I}}})$
			\STATE Sample $\alpha_{t}=1-\beta_{t}$ and $\bar{\alpha}_{t}={\textstyle\prod_{i=1}^{t}\alpha_{i}}$
			\FOR{$i=S$ to $1$}
			\STATE $t=(i-1)\cdot T/S+1$
			\STATE $t-1=(i-2)\cdot T/S+1$ if $i>1$, else 0
			\STATE $
			\mathrm{\hat{x}}^{l}_{t-1}=\frac{1}{\sqrt{\alpha_{t}}}(\mathrm{\hat{x}}_{t}^{l}-\frac{(1-{\alpha}_{t})}{\sqrt{1-\bar{\alpha}_{t}}}\epsilon_{\theta}(\mathrm{\hat{x}}^{l}_{t},\mathrm{x}^{l}_{d},\mathrm{\hat{x}}^{h}_{d},t))$\\
			\hspace*{0.7cm}  $+\frac{(1-{\alpha}_{t})(1-\bar{\alpha}_{t-1})}{1-\bar{\alpha}_{t}}$
			\STATE \textit{\# Global color correction}
			\STATE $\mathrm{\hat{x}}^{l}_{t-1}=GCC(\mathrm{\hat{x}}^{l}_{t-1})$
			\ENDFOR
			\STATE Perform a single gradient descent step on
			\STATE $\nabla_{\theta}\|\mathrm{\hat{x}}_{d}^{l}-\mathrm{x}_{0}^{l}\|^{2}$ and $\nabla_{\theta}\|\mathrm{\hat{x}}_{d}^{h}-\mathrm{x}_{0}^{h}\|^{2}$
			%\STATE \textit{\# Inverse discrete wavelet transform}
			%\STATE $I_{out}=IDWT(\{\mathrm{\hat{x}}_{0}^{l},\mathrm{\hat{x}}^{h}_{d}\})$
		\end{algorithmic}
		\textbf{Until} converged\\
		%\textbf{Output}: The enhanced image $I_{out}$.
	\end{algorithm}

	In addition, we define a content loss function $\mathcal{L}_{content}$ to ensure the content consistency between the restored low-frequency component $\mathrm{\hat{x}}_{d}^{l}$ and target sample $\mathrm{x}_{0}^{l}$, accomplished through the minimization of their $L_{1}$ distance and structural similarity. This content-based constraint function $\mathcal{L}_{content}$ is customized as follows:
	\begin{equation}\label{Content_loss_function}
	\mathcal{L}_{content}=\left\|\mathrm{\hat{x}}_{d}^{l}-\mathrm{x}_{0}^{l}\right\|_{1}+(1-SSIM(\mathrm{\hat{x}}_{d}^{l},\mathrm{x}_{0}^{l})).
	\end{equation}
	
	\begin{algorithm}[!ht]
		\caption{The Inference of DiffColor}
		\label{Sampling}
		\textbf{Input}: Degraded image $\mathrm{X}_{d}$; Noise schedule $\beta_{t}$; Noise estimation network $\theta$; Diffusion steps $T$ and sampling steps $S$.\\
		%\vspace{-\baselineskip}
		\begin{algorithmic}[1]
			\STATE \textit{\# Discrete wavelet transform}
			\STATE $\{\mathrm{x}^{l}_{d},\mathrm{x}^{h}_{d}\}=DWT(\mathrm{X}_{d})$
			\STATE \textit{\# Refine high-frequency components}
			\STATE
			$\mathrm{\hat{x}}^{h}_{d}=CSDR(\mathrm{{x}}^{h}_{d})$ %$\{\hat{V}^{k}_0,\hat{H}^{k}_0,\hat{D}^{k}_0\}=CSDR(\{V^{k}_d,H^{k}_d,D^{k}_d\})$
			\STATE \textit{\# Reverse denoising process}
			\STATE Sample ${\mathrm{x}}_{T}\sim \mathcal N(0,\mathrm{\textit{\textbf{I}}})$
			\STATE Sample $\alpha_{t}=1-\beta_{t}$ and $\bar{\alpha}_{t}={\textstyle\prod_{i=1}^{t}\alpha_{i}}$
			\FOR{$i=S$ to $1$}
			\STATE $t=(i-1)\cdot T/S+1$
			\STATE $t-1=(i-2)\cdot T/S+1$ if $i>1$, else 0
			\STATE $
			\mathrm{\hat{x}}^{l}_{t-1}=\frac{1}{\sqrt{\alpha_{t}}}(\mathrm{\hat{x}}_{t}^{l}-\frac{(1-{\alpha}_{t})}{\sqrt{1-\bar{\alpha}_{t}}}\epsilon_{\theta}(\mathrm{\hat{x}}^{l}_{t},\mathrm{x}^{l}_{d},\mathrm{\hat{x}}^{h}_{d},t))$\\
			\hspace*{0.7cm}  $+\frac{(1-{\alpha}_{t})(1-\bar{\alpha}_{t-1})}{1-\bar{\alpha}_{t}}$
			\STATE \textit{\# Global color correction}
			\STATE $\mathrm{\hat{x}}^{l}_{t-1}=GCC(\mathrm{\hat{x}}^{l}_{t-1})$
			\ENDFOR
			\STATE \textit{\# Inverse discrete wavelet transform}
			\STATE $I_{out}=IDWT(\mathrm{\hat{x}}_{0}^{l},\mathrm{\hat{x}}^{h}_{d})$
			\STATE\textbf{Return $I_{out}$} 
		\end{algorithmic}
	\end{algorithm}
	
	Based on the noise estimation term $\mathcal{L}_{noise}$, the details refinement term $\mathcal{L}_{details}$, and the content preservation term $\mathcal{L}_{content}$, the total objective function $\mathcal{L}_{total}$ is defined by combining them as follows:
	\begin{equation}\label{Total_loss_function}
	\mathcal{L}_{Total}=\mathcal{L}_{noise}+\lambda\mathcal{L}_{details}+\mathcal{L}_{content},
	\end{equation}
	where $\lambda=0.1$ is weighted to coordinate the significance of details refinement with the other two terms in the experiment.
	
	By transforming the input image from the image domain to the wavelet domain, its spatial dimension is reduced, thereby improving the restoration efficiency while leveraging the powerful generation capability of the diffusion model. The enriched three high-frequency spectra $\mathrm{\hat{x}}_{0}^{h} \in \{\hat{V}^{k}_\mathrm{0},\hat{H}^{k}_\mathrm{0},\hat{D}^{k}_\mathrm{0}\}$ at scale $k$-th and their corresponding restored low-frequency component $\mathrm{\hat{x}}_{0}^{l}$ are transformed to the final enhanced result by employing the Inverse Discrete Wavelet Transform (IDWT), expressed as follows:
	\begin{equation}\label{IDWT_function}
	I_{out}=IDWT(\mathrm{\hat{x}}_{0}^{l},\{\hat{V}^{k}_\mathrm{0},\hat{H}^{k}_\mathrm{0},\hat{D}^{k}_\mathrm{0}\}),
	\end{equation}
	where $I_{out}$ represents the final restored image. Algorithm \ref{Training} and Algorithms \ref{Sampling} present the specific training and inference processes, respectively. %The training phase contains forward diffusion and denoising that contribute to the model achieving stable sampling, whereas the inference phase exclusively involves the reverse denoising process.
	
	\subsection{Low-frequency Color Correction via Conditional Sequential Modulation}
	Underwater degraded images usually exhibit a wide variety of degradation appearances due to the unbalanced distribution of color channels resulting from the light selective absorption, as shown in Fig. \ref{1-2Various_Degraded_Types}. However, these various intra-domain color shifts cannot be adequately corrected with diffusion models alone, which has been experimentally demonstrated in \cite{Zhou2023PyramidDM}. Guided by the principle \cite{Wang2023UCD} that natural images typically exhibit similar channel distributions, we propose an adaptive color correction method to modulate the three-channel information of the sampled results during the reverse denoising.
	
	%It consists of three Feature Modulation (FM) blocks, three Strided Convolutional layers with a stride of 2, and three Convolutional layers with a size of $1\times 1$.
	
	\begin{figure*}[!ht]
		\setlength{\abovecaptionskip}{0.0cm}
		\setlength{\belowcaptionskip}{-0.2cm}
		\centering
		\includegraphics[width=0.8\textwidth]{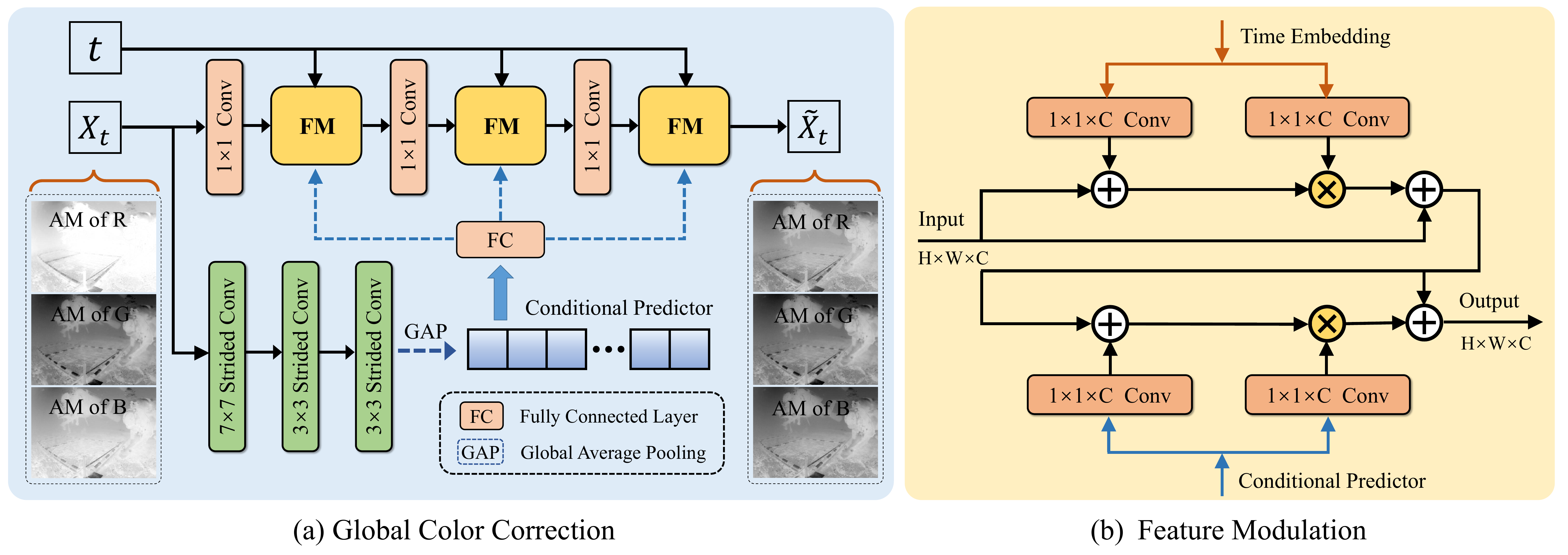}
		\caption	
		{The detailed architecture of the Global Color Correction (GCC) module and Feature Modulation (FM) block. The GCC module consists of a conditional sequence network and a baseline network with feature modulation.
		}
		\label{3-0GCC_FM}
		\vspace{-0.2cm}
	\end{figure*}
	
	As depicted in Fig. \ref{3-0GCC_FM} (a), the global color correction (GCC) module mainly consists of a conditional prediction network and a feature-modulated baseline network. The baseline network employs a fully convolutional structure with a size of $1\times 1$ to ensure that each pixel in the input image is manipulated independently, while the conditional predictor network uses dilated convolution to ensure that the global prior can be propagated to the base network in the manner of feature modulation. The condition network resembles an encoded predictor, comprising three stride convolution blocks for generating a 32-dimensional $1\times 1$ vector from the input image, \textit{i.e.}, conditional predictor. Specifically, the network comprises three Strided Convolutional layers, each with a channel size of 32. The kernel size of the first Strided Convolutional layer is set to $7\times 7$ to increase the receptive field, while the last two layers are $3\times 3$. Each strided convolutional layer sets the stride to 2 and downsamples the feature map to half. At the end of the condition network, the Global Average Pooling (GAP) layer is used to obtain a $1\times1$ 32-dimensional vector. Finally, the predictor generates conditional parameters based on the degradation features of underwater images, which are then propagated in a fully connected manner to the baseline network for image correction.
	
	%\cite{He2019GFM}
	
	The conditional network estimates a global degradation prior from the input underwater image, which is then propagated to the baseline network to assist in correcting the RGB channel information. Given a feature $x$ with the shape of $H\times W\times C$, the instance-level normalization method \cite{jin2020Normalize} is calculated on the feature $x$ by subtracting the mean $\mu(x)$ and then dividing by the standard deviation $\sigma(x)$, which is expressed as follows:
	
	\begin{equation}\label{Global Feature Modulation}
	F_{out}(x) = \gamma\frac{x-\mu(x)}{\sigma(x)}+\beta,
	\end{equation}
	where $\mu(x)$ and $\sigma(x)$ are the mean and standard deviation of the input feature map $x$, and $\gamma, \beta \in \mathbb{R}^{C}$ are scalable parameters learned from data distribution. 
	
	To facilitate global prior estimation, we revised the normalization manner to achieve the color correction of the generated image $X_{t}$ only through scaling and shifting operations. The revised feature modulation (FM) only requires $\gamma$ and $\beta$ to scale and shift the feature map $x$ without normalizing it. Thus, the normalization operation in Eq. (\ref{Global Feature Modulation}) can be modified as $F_{out}(x) = \gamma\ast x+\beta$. Fig. \ref{3-0GCC_FM} (b) illustrates the detailed structure of the FM module, which mainly modulates channel features with these affine parameters generated by the conditional prediction network. Since the gradual reduction of noise during inverse inference is crucial prior information, we ensure that the corrected results establish a certain correspondence with the noise level by adding time embedding $t$, serving as an input to the feature modulation. As illustrated in Fig. \ref{3-0GCC_FM} (a), the channel attenuation map (AM) of the corrected image $\tilde{X}_{t}$ using the GCC module achieves a balanced distribution, whereas the input sampled result $X_{t}$ exhibit significant attenuation.

	\begin{figure}[!ht]
		\setlength{\abovecaptionskip}{0.0cm}
		\setlength{\belowcaptionskip}{-0.2cm}
		\centering
		\includegraphics[width=0.476\textwidth]{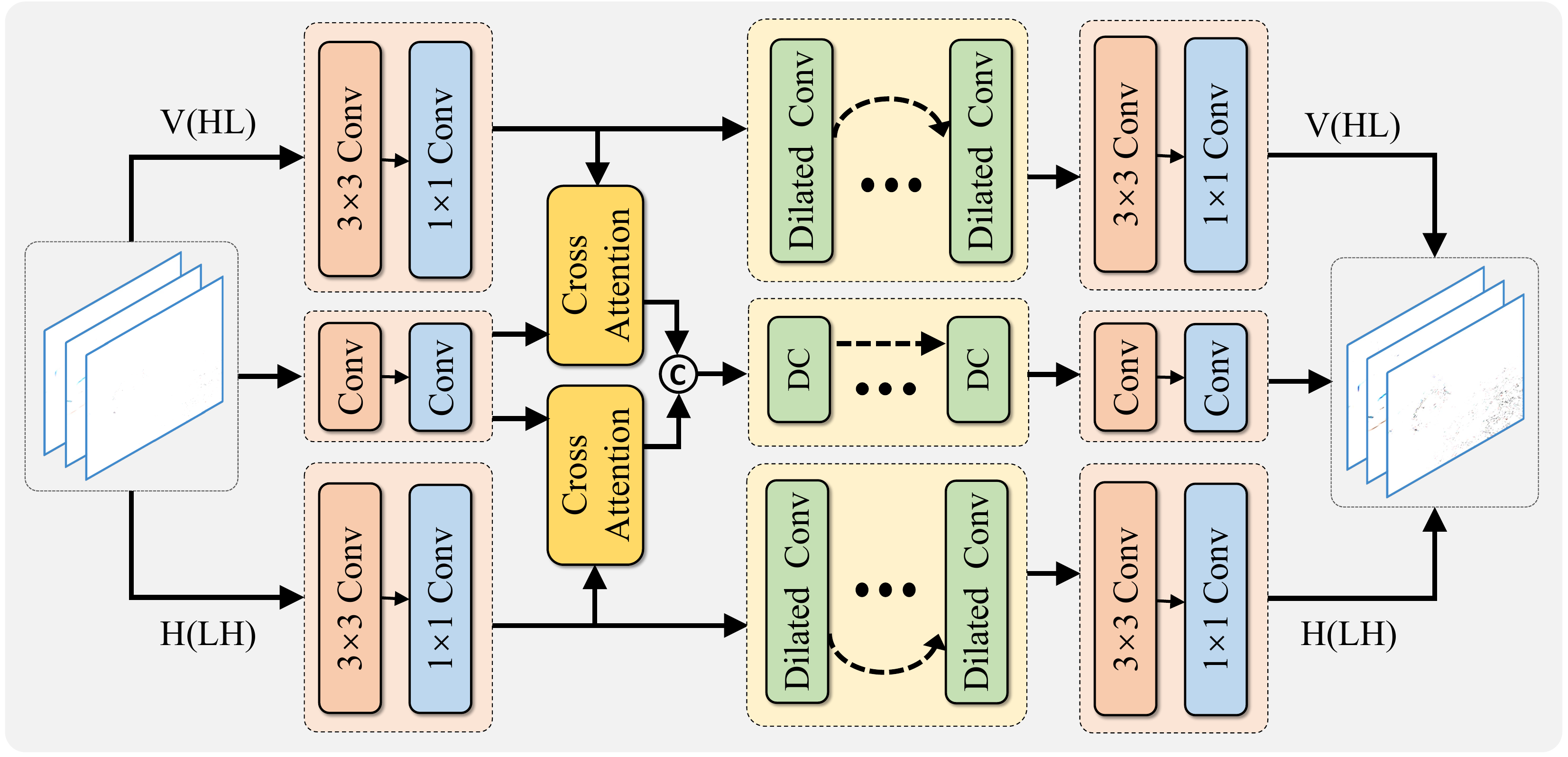}
		\caption
		{The detailed architecture of the Cross-spectral Detail Refinement module (CSDR). It includes six groups of $3\times 3$ and $1\times 1$ convolutional layers, three dilated convolutional blocks, and two cross-attention layers that compensate for diagonal information by relying on vertical and horizontal high-frequency components.}
		\label{3-3Cross-spectral_Detail_Refinement}
		\vspace{-0.2cm}
	\end{figure}
	
	\subsection{High-frequency Details Refinement via Cross-spectral Content Collaboration}\label{High-frequency_Details_Refinement}
	Underwater imaging is affected by medium scattering, resulting in blurred image details and even sacrificing many crucial pieces of information. However, many previous methods have focused on correcting the color cast in enhancing underwater images, and tend to neglect the high-frequency refinement that carries more texture information. To remedy their shortcomings, we introduce a Cross-spectral Detail Refinement (CSDR) to enhance the sacrificed image details.
	
	\begin{figure*}[!htp]
		\setlength{\abovecaptionskip}{0.0cm}
		\setlength{\belowcaptionskip}{-0.2cm}
		\centering
		\includegraphics[width=1.0\textwidth]{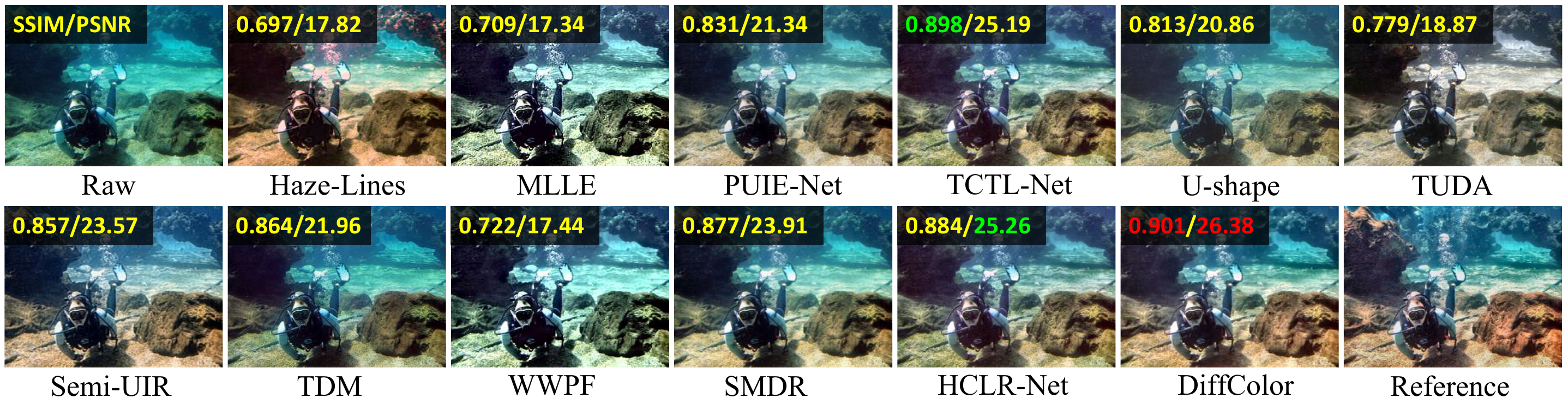}
		\caption	
		{Qualitative results of testing images from the full-reference underwater benchmark \textbf{\textit{Test-U100}}. From left to right: Input image, Haze-Lines \cite{Berman2021HLSQUID}, MLLE \cite{Zhang2022MLLE}, PUIE-Net \cite{Fu2022PUIE-Net}, TCTL-Net \cite{Li2023TCTL-Net}, U-shape \cite{Peng2023U-shapeLSUI}, TUDA \cite{Wang2023TUDA}, Semi-UIR \cite{Huang2023Semi-UIR}, TDM \cite{Tang2023DM-based}, WWPF \cite{Zhang2024WWPF}, SMDR \cite{Zhang2024SMDR}, HCLR-Net \cite{Zhou2024HCLR}, our DiffColor, and the clear reference.}
		\label{5-1With_ReferenceU100}
		\vspace{-0.2cm}
	\end{figure*}
	\begin{figure*}[!htp]
		\setlength{\abovecaptionskip}{0.0cm}
		\setlength{\belowcaptionskip}{-0.2cm}
		\centering
		\includegraphics[width=1.0\textwidth]{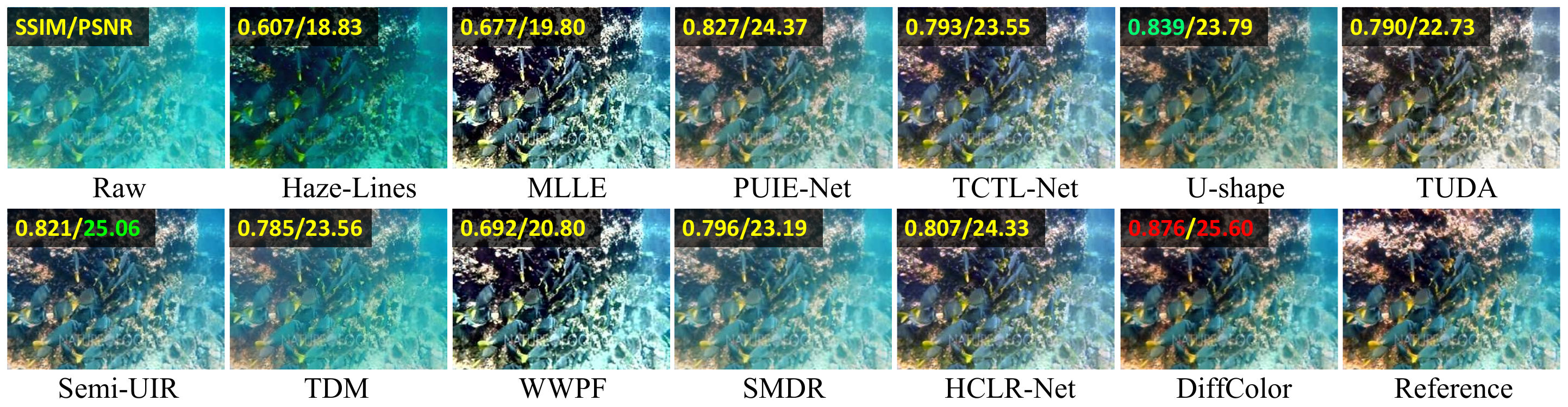}
		\caption	
		{Qualitative results of testing images from the full-reference underwater benchmark \textbf{\textit{Test-L500}}. From left to right: Input image, Haze-Lines \cite{Berman2021HLSQUID}, MLLE \cite{Zhang2022MLLE}, PUIE-Net \cite{Fu2022PUIE-Net}, TCTL-Net \cite{Li2023TCTL-Net}, U-shape \cite{Peng2023U-shapeLSUI}, TUDA \cite{Wang2023TUDA}, Semi-UIR \cite{Huang2023Semi-UIR}, TDM \cite{Tang2023DM-based}, WWPF \cite{Zhang2024WWPF}, SMDR \cite{Zhang2024SMDR}, HCLR-Net \cite{Zhou2024HCLR}, our DiffColor, and the clear reference.}
		\label{5-2With_ReferenceL500}
		%\vspace{-0.2cm}
	\end{figure*}

	The discrete wavelet transform divides the high-frequency spectrum of an image into vertical, horizontal, and diagonal components, each of which exhibits complementary interdependencies. Benefiting from this property, we design the CSDR module that leverages the vertical and horizontal information to compensate for the sacrificed details in the diagonal. As shown in Fig. \ref{3-3Cross-spectral_Detail_Refinement}, the three high-frequency components $\{V^{k}_d,H^{k}_d,D^{k}_d\}$ are sequentially extracted by three $3\times3$ and $1\times1$ Convolutional layers, and then two Cross-attention layers \cite{Hou2019CrossAttention} are employed to utilize vertical $\{V^{k}_d\}$ and horizontal $\{H^{k}_d\}$ information to restore the details in diagonal $\{D^{k}_d\}$. Specifically, taking the high-frequency components $\{V^{k}_d\}$ and $\{D^{k}_d\}$ as examples, we calculate the corresponding query vector $Q_{vd}=query(V^{k}_d)$, key vector $K_{vd}=key(D^{k}_d)$, and value vector $V_{vd}=key(D^{k}_d)$, respectively. Then, the first output feature $F_{VD}$ of the cross attention block are calculated as follows:
	\begin{equation}\label{Cross_attention_function}
	F_{VD}=Softmax\left(\frac{Q_{vd}K_{vd}^{T}}{\sqrt{d_{x}}}V_{vd}\right),
	\end{equation}
	where $d_{x}$ denotes the dimension of the key vector $K_{vd}$. Similarly, we can calculate the corresponding cross-attention feature result $F_{HD}$ based on the horizontal and diagonal high-frequency components in the same manner.
	%Motivated by this design \cite{Jiang2023LLDM}, 
	
	To increase the receptive field of the input, we adopt the Dilated Convolutional blocks with a progressive dilation rate for feature extraction, in which the first and last layers are used to extract local information, and the middle layers improve global information utilization by expanding the receptive field. By setting the dilation rates as $d=\{1,2,3,2,1\}$, this gradual increase and decrease effectively avoids grid effects. Finally, three refined high-frequency components $\{\hat{V}^{k}_d,\hat{H}^{k}_d,\hat{D}^{k}_d\}$ are obtained by reducing the channel dimensions with three $3\times3$ and $1\times1$ Convolutional layers. As a whole, the CSDR module sensibly leverages the complementary advantages of the three high-frequency components $\{V^{k}_d, H^{k}_d, D^{k}_d\}$ to refine the textural details of degraded images.

	\section{Experiment and Analysis}
	\subsection{Experimental Settings}
	\textbf{Implementation Details.} The proposed DiffColor is trained using the Pytorch framework on two NVIDIA GeForce RTX 4090 GPUs for 500 epochs. During the training phase, the batch size and patch size are set to 16 and $256\times256$, respectively. The Adam optimizer comes with an initial learning rate of $1\times10^{-4}$ and decreases it by a factor of 0.8 after every fifty epochs. The diffusion time step is set to $T=200$, while the sampling step is set to $S=10$ for efficient restoration. The number of wavelet transforms is set to 2, and the resolution of the low-frequency component is reduced to 1/4 of the original input image.
	
	\textbf{Underwater Image Datasets.} Our experiments include paired image benchmarks (UIEB \cite{Li2020UIEB}, EUVP \cite{Islam2020EUVP}, and LSUI \cite{Peng2023U-shapeLSUI}), consisting of real-world underwater images with ground truth references. The benchmarks of RUIE \cite{Liu2020RUIE}, SQUID \cite{Berman2021HLSQUID}, and UDD \cite{Liu2021UDD} lack paired images and instead comprise degraded single underwater images. Specifically, following \cite{Tang2023DM-based, Wang2023TUDA}, we utilize 790 images from the UIEB dataset and 3779 images from the LSUI dataset for training. The remaining 100 and 500 underwater images in UIEB and LSUI are designated as Test-U100 and Test-L500 for testing, respectively. During testing, images required for the full-reference evaluation primarily originate from EUVP, Test-U100, and Test-L500, while the non-reference evaluation relies on the benchmarks of RUIE, SQUID, and UDD. %All images are scaled to a fixed size of $256\ast 256$ as they are fed into the network and the pixel values are normalized to between $\left [0, 1\right ]$.
	
	\begin{figure*}[!htp]
		\setlength{\abovecaptionskip}{0.0cm}
		\setlength{\belowcaptionskip}{-0.2cm}
		\centering
		\includegraphics[width=1.0\textwidth]{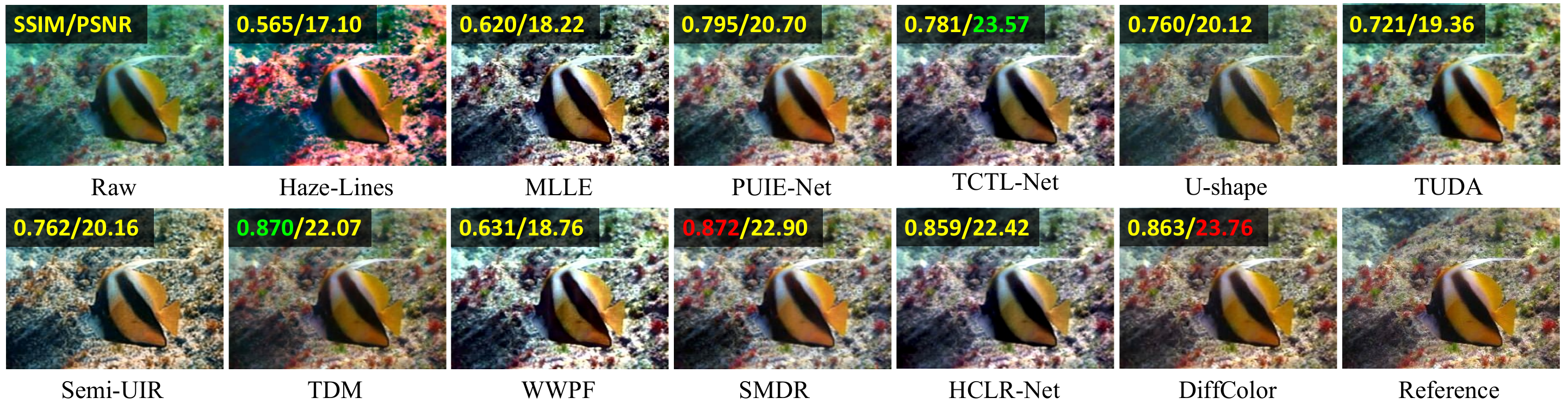}
		\caption	
		{Qualitative results of testing images from the full-reference underwater benchmark \textbf{\textit{EUVP}} \cite{Islam2020EUVP}. From left to right: Input image, Haze-Lines \cite{Berman2021HLSQUID}, MLLE \cite{Zhang2022MLLE}, PUIE-Net \cite{Fu2022PUIE-Net}, TCTL-Net \cite{Li2023TCTL-Net}, U-shape \cite{Peng2023U-shapeLSUI}, TUDA \cite{Wang2023TUDA}, Semi-UIR \cite{Huang2023Semi-UIR}, TDM \cite{Tang2023DM-based}, WWPF \cite{Zhang2024WWPF}, SMDR \cite{Zhang2024SMDR}, HCLR-Net \cite{Zhou2024HCLR}, our DiffColor, and the clear reference.}%The best evaluation scores for each method in terms of SSIM/PSNR metrics are highlighted in red, while the second-best scores are highlighted in blue.
		\label{5-3With_ReferenceEUVP}
		\vspace{-0.2cm}
	\end{figure*}
	\begin{figure}[!htp]
		\setlength{\abovecaptionskip}{0.0cm}
		\setlength{\belowcaptionskip}{-0.3cm}
		\centering
		\includegraphics[width=0.47\textwidth]{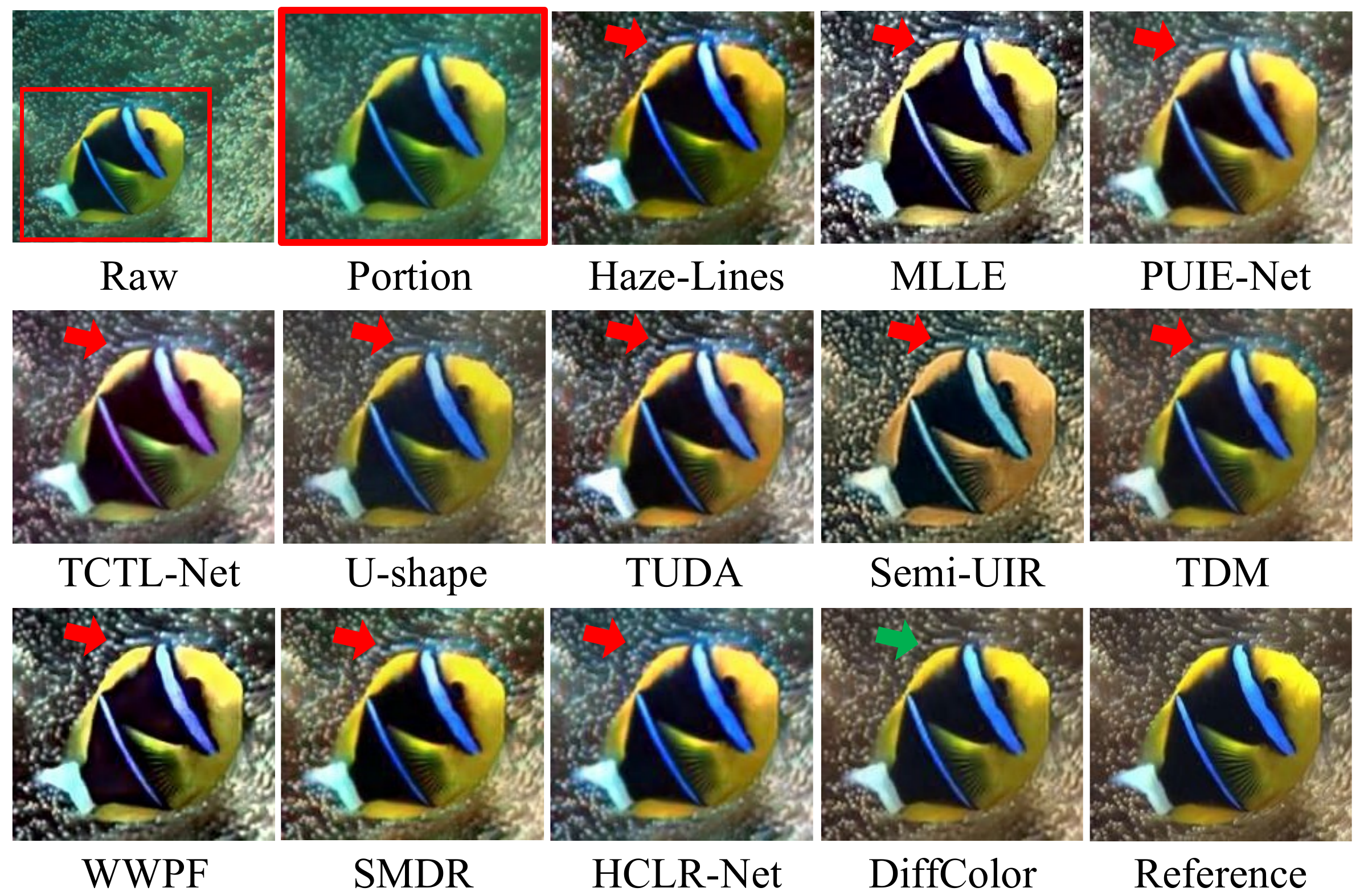}
		\caption
		{Qualitative comparison of color and details preservation. DiffColor achieves the most homogeneous visual perception and does not introduce blue-green artifacts around the fish, as observed in other methods (marked with red arrows).}
		\label{5-5Detail_Preserving}
	\end{figure}

	\textbf{Comparison with the SOTA Methods.} We compare the proposed DiffColor with eleven state-of-the-art (SOTA) underwater image enhancement methods, including Haze-Lines \cite{Berman2021HLSQUID}, MLLE \cite{Zhang2022MLLE}, PUIE-Net \cite{Fu2022PUIE-Net}, TCTL-Net \cite{Li2023TCTL-Net}, U-shape \cite{Peng2023U-shapeLSUI}, TUDA \cite{Wang2023TUDA}, Semi-UIR \cite{Huang2023Semi-UIR}, TDM \cite{Tang2023DM-based}, WWPF \cite{Zhang2024WWPF}, SMDR \cite{Zhang2024SMDR}, and HCLR-Net \cite{Zhou2024HCLR}. Among these methods, Haze-Lines, MLLE, and WWPF are three physical-based methods, while PUIE-Net, TCTL-Net,  U-shape, TUDA, Semi-UIR, TDM, SMDR, and HCLR-Net are seven deep learning-based methods.

	%\begin{figure*}[!htp]
	%	\setlength{\abovecaptionskip}{0.1cm}
	%	%\setlength{\belowcaptionskip}{-0.2cm}
	%	\centering
	%	\includegraphics[width=0.986\textwidth]{Figs/5-5Detail_Preserving.pdf}
	%	\caption	
	%	{Qualitative comparison of details preservation and artifact removal. Our method achieves the closest perception to the reference and does not introduce blue-green artifacts around the fish (marked by a red arrow) like others.}
	%	\label{5-5Detail_Preserving}
	%\vspace{-0.1cm}
	%\end{figure*}
	
	\textbf{Evaluation Metrics.} For paired images, we adopt three full-reference metrics to evaluate the enhancement performance of each method, including UIF \cite{Zheng2022UIF}, SSIM, and PSNR. Meanwhile, for these unpaired images, the results are evaluated with UCIQE \cite{Yang2015UCIQE} and UIQM \cite{Panetta2016UIQM}, which are popular non-reference metrics for evaluating underwater images. Specifically, UIF utilizes statistical features in CIELab space to evaluate underwater images in terms of naturalness, clarity, and structural indexes. The UCIQE metric is used to evaluate the quality of color distribution in the Lab color space. The UIQM consists of three attributes called UICM, UISM, and UIConM, which are used to evaluate the color, clarity, and contrast of underwater image perception, respectively. Throughout the experiment, we maintain the parameters suggested in the respective papers for each metric.

	\subsection{Qualitative Evaluation with Naturalness}
	We first compare the proposed DiffColor and ten SOTA methods on three full-reference underwater datasets, including Test-U100, Test-L500, and EUVP. The enhancement results on three underwater datasets are showcased in Fig. \ref{5-1With_ReferenceU100}, Fig. \ref{5-2With_ReferenceL500}, and Fig. \ref{5-3With_ReferenceEUVP}, respectively. In the visual comparison, even though most methods have achieved satisfactory enhancement results, our method demonstrates the closest perception to the reference in terms of color saturation, content similarity, and global contrast. Considering the subjectivity of individual visual perception, we further employ two commonly used image quality evaluation metrics (SSIM and PSNR) to compare the enhancement results of each method. As shown in Fig. \ref{5-1With_ReferenceU100}-\ref{5-3With_ReferenceEUVP}, the results show that our method achieves the best scores on all three datasets of Test-U100, Test-L500, and EUVP. Semi-UIR and HCLR-Net achieved the second-best scores on both test-U100 and Test-L500 datasets, while TUDA and TDM achieved relatively satisfactory results on the sample image from the EUVP benchmark. To verify the color fidelity of each method in handling underwater images, we conducted a detailed comparison with respect to color preservation. As shown in Fig. \ref{5-5Detail_Preserving}, our DiffColor achieves the most homogeneous visual perception and does not introduce blue-green artifacts around the fish as observed in other methods. Overall, our DiffColor surpasses several state-of-the-art underwater image enhancement methods on reference datasets.

	%In real-world underwater scenarios, there is a lack of high-fidelity clear images to serve as references for degraded images. We conduct a comparison on three non-reference benchmarks RUIE, SQUID, and UDD, and the qualitative results are shown in Fig. \ref{5-4With_Non-Reference}. When confronted with various degradation scenes, our DiffColor has demonstrated superior robustness in terms of detail preservation and color restoration compared to other methods. To verify the color fidelity of each method in handling underwater images, we conducted a detailed comparison regarding color preservation on standard color charts. Fig. \ref{5-6Detail_Preserving_With_Reference} shows the detailed comparison with the standard color chart for reference. Therefore, various qualitative results validated that our DiffColor outperforms other underwater image enhancement methods in equalizing the brightness, contrast, and color fidelity. 
	
	\subsection{Quantitative Evaluation with Metrics}
	The results of quantitative evaluation in Test-U100, Test-L500, and EUVP are presented in Table \ref{Full_reference_Metrics}. Our proposed DiffColor achieves the best results in terms of UIF, SSIM, and PSNR metrics. The potential reason is the robust generative capability of the designed diffusion model, coupled with the CSDR and GCC modules that focus on image detail richness and color correction. For the SSIM metric, HCLR-Net, U-shape, and TDM methods achieve the second-best results on Test-U100, Test-L500, and EUVP benchmarks, respectively. In addition, Semi-UIR and PUIE-Net methods obtain the second-best scores on Test-U100 and EUVP datasets in terms of PSNR metrics, respectively.
	
	\begin{table*}
		\setlength{\abovecaptionskip}{0.0cm}
		\setlength{\belowcaptionskip}{-0.2cm}
		\renewcommand\arraystretch{1.1}
		\tabcolsep 0.096 in
		\caption{Quantitative evaluation of full-reference metrics on three underwater datasets (\textbf{\textit{Test-U100}}, \textbf{\textit{Test-L500}}, and \textbf{\textit{EUVP}}). The best and second-best results are highlighted with \textbf{bold} and \underline{underlined}, respectively.}
		\label{Full_reference_Metrics}
		\centering
		\begin{tabular}{c|ccc|ccc|ccc|ccc}
			\hline
			\hline
			\multirow{2}{*}{\textbf{Methods}}&\multicolumn{3}{c|}{\textbf{Test-U100}}&\multicolumn{3}{c|}{\textbf{Test-L500}}&\multicolumn{3}{c|}{\textbf{EUVP}}&\multicolumn{3}{c}{\textbf{Average}}\\
			\cline{2-13} &UIF$\uparrow$&SSIM$\uparrow$&PSNR$\uparrow$&UIF$\uparrow$&SSIM$\uparrow$&PSNR$\uparrow$&UIF$\uparrow$&SSIM$\downarrow$&PSNR$\downarrow$&UIF$\uparrow$&SSIM$\downarrow$&PSNR$\downarrow$\\
			\hline
			Haze-Lines \cite{Berman2021HLSQUID} &0.435	&0.616	&16.376	&0.416	&0.551	&19.786	&0.343	&0.563	&18.652	&0.359 	&0.563 	&18.774 \\
			MLLE \cite{Zhang2022MLLE} &0.405 &0.631 &16.660 &0.365 	&0.632 	&19.202 &0.347 	&0.584	&17.268 &0.352 &0.594 &17.593\\
			PUIE-Net \cite{Fu2022PUIE-Net} &0.486 &0.825 &21.841 &0.454 &0.821 &24.354 &\underline{0.417} &0.802 &22.778 &\underline{0.426} &0.806 &23.027 \\
			TCTL-Net \cite{Li2023TCTL-Net}	&\textbf{0.512} &\textbf{0.906}	&\underline{25.357} &0.445 &0.788 &22.830 &0.407 &0.736 &21.781 &0.418 &0.751 &22.098 \\
			U-shape \cite{Peng2023U-shapeLSUI} &0.476 &0.816 &21.826 &0.446 &0.849 &\underline{24.631} &0.410 &0.819 &\underline{23.392} &0.419 &0.824 &\underline{23.558} \\
			TUDA \cite{Wang2023TUDA} &0.491 &0.787 &18.239 &0.441 &0.793 &21.703 &0.407 &0.723 &20.277 &0.416 &0.738 &20.460\\
			Semi-UIR \cite{Huang2023Semi-UIR} &0.493 &0.852 &23.636 &0.433 &0.815 &24.070 &0.394 &0.796 &22.107 &0.405 &0.801 &22.514		\\
			TDM \cite{Tang2023DM-based} &0.454 &0.876 &21.657 &\textbf{0.463} &0.793 &24.470 &0.408 &\textbf{0.889} &23.277 &0.420 &0.871 &23.433 \\
			WWPF \cite{Zhang2024WWPF} &0.446 &0.711 &17.571 &0.389 &0.680 &19.419 &0.361 &0.642 &18.038 &0.369 &0.651 &18.269\\
			SMDR \cite{Zhang2024SMDR} &0.496 &0.870 &23.575 &0.448 &\underline{0.855} &24.219 &0.411 &\underline{0.880} &22.647 &0.421 &\underline{0.875} &22.963 \\
			HCLR-Net \cite{Zhou2024HCLR} &0.491 &0.882 &25.174 &0.449 &0.836 &23.656 &0.413 &0.856 &22.747 &0.422 &0.853 &22.997\\
			DiffColor &\underline{0.498} &\underline{0.891} &\textbf{26.165} &\underline{0.461} &\textbf{0.861} &\textbf{25.636} &\textbf{0.433} &0.879	&\textbf{23.518}	&\textbf{0.440}	&\textbf{0.876}	&\textbf{23.993} \\
			\hline
			\hline
		\end{tabular}
		\vspace{-0.2cm}
	\end{table*}
	%\begin{table}[!htp]
		%\setlength{\abovecaptionskip}{0.1cm}
		%\renewcommand\arraystretch{1.1}
		%\tabcolsep 0.127 in
		%\caption{Efficiency of each method with Parameters (M), FLOPs (G), Inference Time (s), and FPS. The best and second-best scores are highlighted with \textbf{bold} and \underline{underlined}, respectively.} %The left is the UIE method for visual restoration, and the right is the USOU method for saliency detection.
		%\label{Parameters_FLOPs_Time}
		%\centering
		%\begin{tabular}{c|cc|cc}
			%\hline
			%\hline
			%Method &Param. $\downarrow$ &FLOPs $\downarrow$ &Time $\downarrow$ &FPS $\uparrow$\\
			%\cline{1-5}
			%Haze-Lines &\textit{Null} &\textit{Null} &2.316 &0.43 \\
			%MLLE &\textit{Null} &\textit{Null} &0.079 &12.68\\
			%PUIE-Net &10.69 &\underline{30.09} &0.203 &4.93\\
			%TCTL-Net &99.72 &56.51 &0.067  &14.93\\
			%U-shape &65.60 &66.20 &0.106 &9.40 \\
			%TUDA &31.36 &174.6 &0.157 &6.37 \\
			%Semi-UIR &12.78 &36.46 &0.631 &1.58 \\
			%TDM &\underline{10.71} &66.89 &0.730 &1.37 \\
			%WWPF &\textit{Null} &\textit{Null} &0.390 &2.56\\
			%SMDR &12.26 &46.59 &\underline{0.062} &\underline{16.13} \\			
			%HCLR-Net &\textbf{4.87} &401.97 &0.276 &3.62 \\
			%DiffColor &22.13 &\textbf{21.99} &\textbf{0.055} &\textbf{18.31} \\
			%\hline
			%\hline
		%\end{tabular}
		%\vspace{-0.3cm}
	%\end{table}
    
	%However, as described in \cite{Wang2023TUDA, Huang2023Semi-UIR}, UIQM and UCIQE tend to favor certain characteristics and thus may not accurately reflect the visual perception of the enhanced results.
	
	\subsection{Evaluation of Model Efficiency}
\begin{figure}[!htp]
		\setlength{\abovecaptionskip}{0.1cm}
		\setlength{\belowcaptionskip}{-0.2cm}
		\centering
		\includegraphics[width=0.456\textwidth]{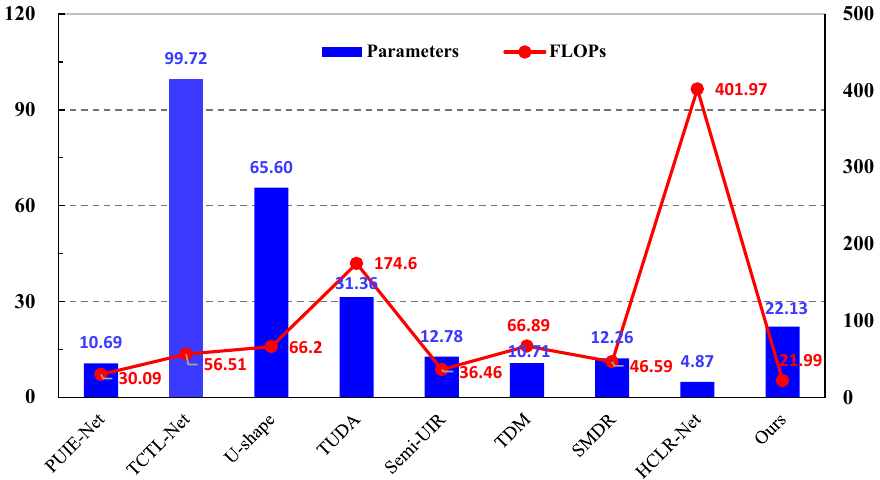}
		\caption
		{Efficiency of each deep learning-based method with Parameters (M) and FLOPs (G).}
		\label{6-2FLOPs_and_Params}
		\vspace{-0.3cm}
	\end{figure}

\begin{figure}[!h]
		\setlength{\abovecaptionskip}{0.1cm}
		\setlength{\belowcaptionskip}{-0.2cm}
		\centering
		\includegraphics[width=0.456\textwidth]{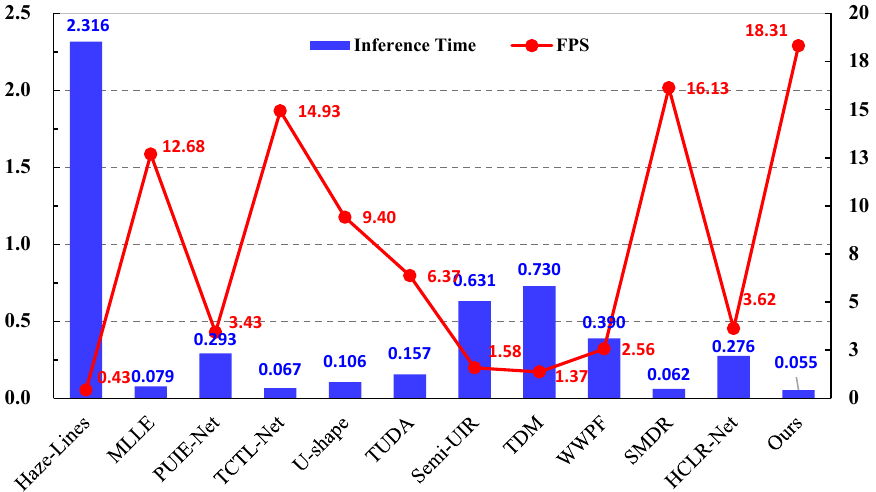}
		\caption
		{Efficiency of each method with Inference Time (s) and FPS.}
		\label{6-1Elapsed_Time}
		\vspace{-0.3cm}
	\end{figure}

	%During the testing phase, all experiments were performed on a PC configured with an Intel Core i7-6500 2.5GHz CPU and 32GB RAM to ensure a fair evaluation of each method. The traditional physics-based methods are run on Matlab 2023a, while the deep learning-based methods are executed on PyCharm 2023 with an NVIDIA GeForce GTX 1080Ti GPU.
	We analyze the model efficiency of DiffColor compared to other methods on Test-U100, Test-L500, and EUVP benchmarks. The comparative experiment covers four aspects: Parameters, FLOPs, inference time, and FPS, with the results depicted in Fig. \ref{6-2FLOPs_and_Params} and Fig. \ref{6-1Elapsed_Time}. All experiments regarding efficiency evaluation are conducted on the same device to ensure fairness across methods.

	\textbf{Parameters and FLOPs Evaluation.}
	%We have provided the inference time and FPS of the proposed DiffColor in Fig. \ref{6-1Elapsed_Time}, which demonstrates the advantage of DiffColor over other SOTA methods. 
	We conduct comparisons between DiffColor and deep learning-based methods in terms of Parameters and FLOPs. As shown in Fig. \ref{6-2FLOPs_and_Params}, we can observe that the FLOPs of DiffColor are the lowest. This is attributed to the introduction of wavelet transform, which reduces the image size, thereby significantly reducing the model's computational overhead. As for the number of parameters, DiffColor is also relatively lightweight compared to TCTL-Net, U-shape, and TUDA methods. Thus, considering both the Params and FLOPs, DiffColor is better suited than other state-of-the-art methods for embedded processors with limited computational power in underwater vehicles.
	 
	%Especially, when applying DiffColor to the real-time processing tasks of underwater videos in real-world applications, the advantages of the proposed method would become more apparent.
	%All experiments regarding method efficiency evaluation were performed on a PC configured with an Intel Core i7-6500 2.5GHz CPU to ensure a fair evaluation of each method. The traditional physics-based methods are run on Matlab 2023a, while the deep learning-based methods are executed based on an NVIDIA GeForce GTX 1080Ti GPU.
	%The advantage of FLOPs also provides support for the improvement of DiffColor in FPS. Our FPS remains the highest among the compared methods in Table \ref{Parameters_FLOPs_Time}.

	\textbf{Inference Time and FPS Evaluation.}
	We further conducted a comparative analysis of the average inference time and Frames Per Second (FPS) of each method. As shown in Fig. \ref{6-1Elapsed_Time}, the proposed DiffColor outperforms both physics-based and deep learning-based state-of-the-art methods. In terms of inference time per image, our DiffColor accelerates over 13.5\% compared to the second fastest SMDR model and surpasses it by 2.19 (18.31 for DiffColor and 16.13 for SMDR) in the FPS metric. Notably, when applied to real-time processing tasks for underwater videos in real-world applications, the advantages of the proposed method are particularly evident.

	%In addition, compared with the similarly diffusion-based TDM method, our DiffColor overwhelmingly reduces the inference time, demonstrating the effectiveness of incorporating discrete wavelet transform into the diffusion models. %In general, our method holds greater promise for achieving real-time enhancement applications in underwater scenarios.
	\begin{figure*}[!h]
		\setlength{\abovecaptionskip}{0.1cm}
		\setlength{\belowcaptionskip}{-0.2cm}
		\centering
		\includegraphics[width=0.86\textwidth]{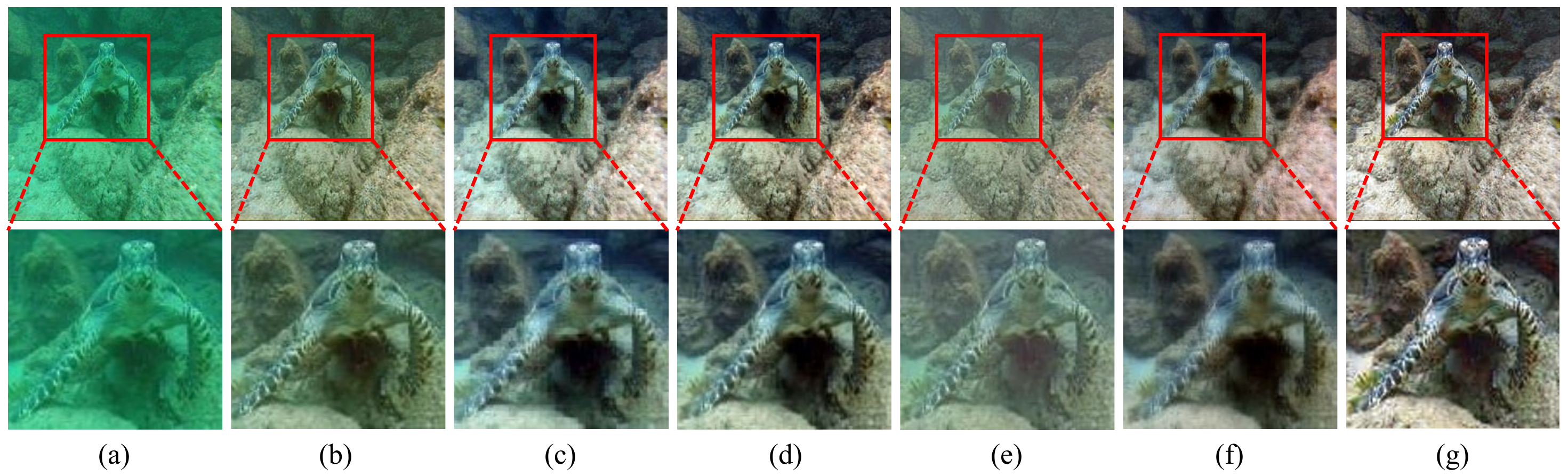}
		\caption	
		{Contribution of each component in the qualitative ablation study. From left to right, (a) Raw underwater image, (b) -w/o GCC: without global color correction, (c) -w/o CSDR: without cross-spectral detail refinement, (d) -w/o \protect$\mathcal{L}_{details}$: without detail-preserving loss, (e) -w/o \protect$\mathcal{L}_{content}$: without content loss, (f) -w/o HF-conditions: without refined high-frequency spectrum as input conditions, and (g) DiffColor: the full framework.}
		\label{6-4Abalation_Qualitative}
		\vspace{-0.2cm}
	\end{figure*}
	\begin{table}[!h]
		\setlength{\abovecaptionskip}{0.0cm}
		\setlength{\belowcaptionskip}{-0.2cm}
		\renewcommand\arraystretch{1.2}
		\tabcolsep 0.036 in
		\caption{Ablation studies of GCC and CSDR modules. ``-w/o'' means there is no corresponding module.}
		\label{Ablation_Studies_on_GCC_and_CSDR_Modules}
		\centering
		\small
		\begin{tabular}{c|cc|cc|cc}
			\hline
			\hline
			\multirow{2}{*}{Datasets}&\multicolumn{2}{c|}{-w/o GCC}&\multicolumn{2}{c|}{-w/o CSDR}&\multicolumn{2}{c}{All}\\
			\cline{2-7}
			&SSIM$\uparrow$&PSNR$\uparrow$&SSIM$\uparrow$&PSNR$\uparrow$&SSIM$\uparrow$&PSNR$\uparrow$\\
			\hline
			Test-U100 &0.783&18.908&0.856&22.780&\textbf{0.891}&\textbf{26.165}\\
			Test-L500 &0.775&20.936&0.832&21.089&\textbf{0.861}&\textbf{25.636}\\
			EUVP &0.801&17.865&0.861&20.106&\textbf{0.879}&\textbf{23.518}\\
			\hline
			\hline
		\end{tabular}
		\vspace{-0.2cm}
	\end{table}
	\begin{table}[!h]
		\setlength{\abovecaptionskip}{0.0cm}
		\setlength{\belowcaptionskip}{0.2cm}
		\renewcommand\arraystretch{1.2}
		\tabcolsep 0.036 in
		\caption{Ablation studies of the loss functions. ``-w/o'' means there is no corresponding loss.}
		\label{Ablation_Studies_on_Loss_Function}
		\centering
		\small
		\begin{tabular}{c|cc|cc|cc}
			\hline
			\hline
			\multirow{2}{*}{Datasets}&\multicolumn{2}{c|}{-w/o $\mathcal{L}_{details}$}&\multicolumn{2}{c|}{-w/o $\mathcal{L}_{content}$}&\multicolumn{2}{c}{All}\\
			\cline{2-7}
			&SSIM$\uparrow$&PSNR$\uparrow$&SSIM$\uparrow$&PSNR$\uparrow$&SSIM$\uparrow$&PSNR$\uparrow$\\
			\hline
			Test-U100 &0.856&22.786&0.817&20.076&\textbf{0.891}&\textbf{26.165}\\
			Test-L500 &0.843&21.341&0.798&19.713&\textbf{0.861}&\textbf{25.636}\\
			EUVP &0.848&21.543&0.809&19.858&\textbf{0.879}&\textbf{23.518}\\
			\hline
			\hline
		\end{tabular}
		\vspace{-0.2cm}
	\end{table}
	
	\begin{table}[!h]
		\setlength{\abovecaptionskip}{0.0cm}
		\setlength{\belowcaptionskip}{0.2cm}
		\renewcommand\arraystretch{1.2}
		\tabcolsep 0.120 in
		\caption{Ablation studies of the conditional inputs.}
		\label{Ablation_Studies_on_Condition}
		\centering
		\small
		\begin{tabular}{c|cc|cc}
			\hline
			\hline
			\multirow{2}{*}{Datasets}&\multicolumn{2}{c|}{-w/o HF-Conditions}&\multicolumn{2}{c}{All}\\
			\cline{2-5}
			&SSIM$\uparrow$&PSNR$\uparrow$&SSIM$\uparrow$&PSNR$\uparrow$\\
			\hline
			Test-U100 &0.829 &23.018&\textbf{0.891}&\textbf{26.165}\\
			Test-L500 &0.816 &22.422&\textbf{0.861}&\textbf{25.636}\\
			EUVP &0.831	&22.205	&\textbf{0.879}&\textbf{23.518}\\
			\hline
			\hline
		\end{tabular}
	\end{table}
	\subsection{Ablation Study}
	To analyze the effectiveness of each component in our DiffColor method, we also conduct an ablation study on Test-U100, Test-L500, and EUVP benchmarks. Fig \ref{6-4Abalation_Qualitative} shows the qualitative results of the ablation contribution for each key component. These results demonstrate that the absence of high-frequency detail refinement leads to the blurring of details in Fig. \ref{6-4Abalation_Qualitative} (c)(f), while without color correction fails to entirely remove the color cast in Fig. \ref{6-4Abalation_Qualitative} (b). In addition, Fig. \ref{6-4Abalation_Qualitative} (d)(e) demonstrates the effectiveness of the loss functions $\mathcal{L}_{details}$ and $\mathcal{L}_{content}$, especially the latter is necessary to remove the foggy appearance of underwater images.
	
	%In addition, Table \ref{Ablation_Studies_on_GCC_and_CSDR_Modules}, Table \ref{Ablation_Studies_on_Loss_Function}, and Table \ref{Ablation_Studies_on_Condition} show the quantitative scores in terms of whether the restoration modules, loss functions, and high-frequency conditions are included, respectively. Fig. \ref{Wavelet_Transform_Levels_and_Sampling_Step} illustrates the ablation studies of wavelet transform levels $K$ and sampling steps $S$ based on three full-reference metrics and inference time.
	
	\textbf{Effectiveness of GCC and CSDR Modules}. In Table \ref{Ablation_Studies_on_GCC_and_CSDR_Modules}, ``-w/o GCC'' means removing the global color correction module, while ``-w/o CSDR'' indicates that there is no cross-spectral detail refinement module to enrich image details. Compared with the whole DiffColor, their SSIM and PSNR scores are degraded on the Test-U100, Test-L500, and EUVP benchmarks, especially for the column of ``-w/o GCC''. The decreased scores illustrate the necessity of color correction and detail refinement throughout the restoration process.
	
	\textbf{Loss Functions}. In Table \ref{Ablation_Studies_on_Loss_Function}, the ``-w/o $\mathcal{L}_{details}$'' denotes the removal of detail-preserving loss $\mathcal{L}_{details}$, which aims to ensure rich details of the restored image. Meanwhile ``-w/o $\mathcal{L}_{content}$'' indicates the removal of the content loss $\mathcal{L}_{content}$, leading to a noticeable decrease in both SSIM and PSNR metrics. This degradation is relatively significant compared to the removal of $\mathcal{L}_{details}$, which highlights the effectiveness of content-preserving loss in our training strategy.
    \begin{figure}[!hbp]
		\centering
		%\subfloat{\label{SSIM_Metric}}\addtocounter{subfigure}{-1}
		\subfloat[SSIM Scores]{
			{\includegraphics[scale=0.276]{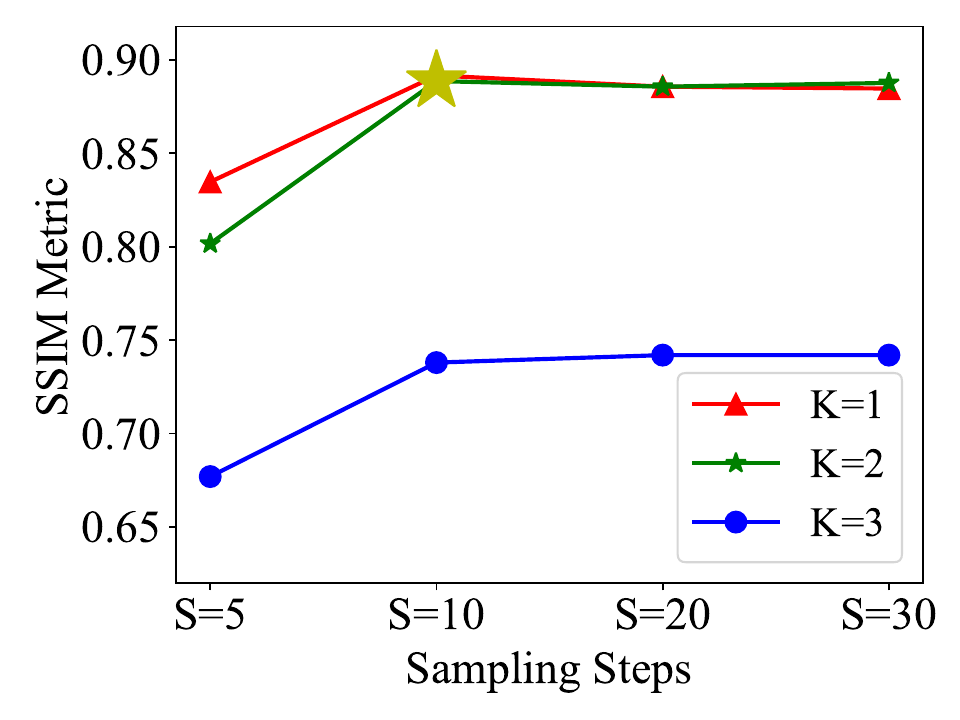}}
			\label{SSIM_Metric}}
		\subfloat[PSNR Scores]{
			{\includegraphics[scale=0.276]{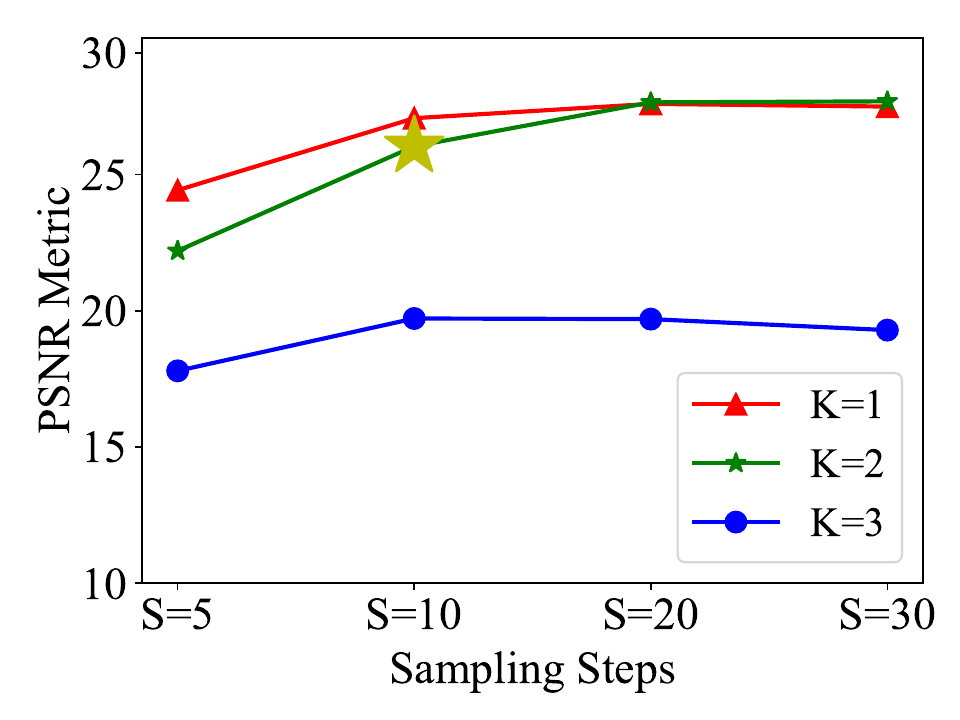}}
			\label{PSNR_Metric}}
		\hfil	
		\subfloat[UIF Scores]{
			{\includegraphics[scale=0.276]{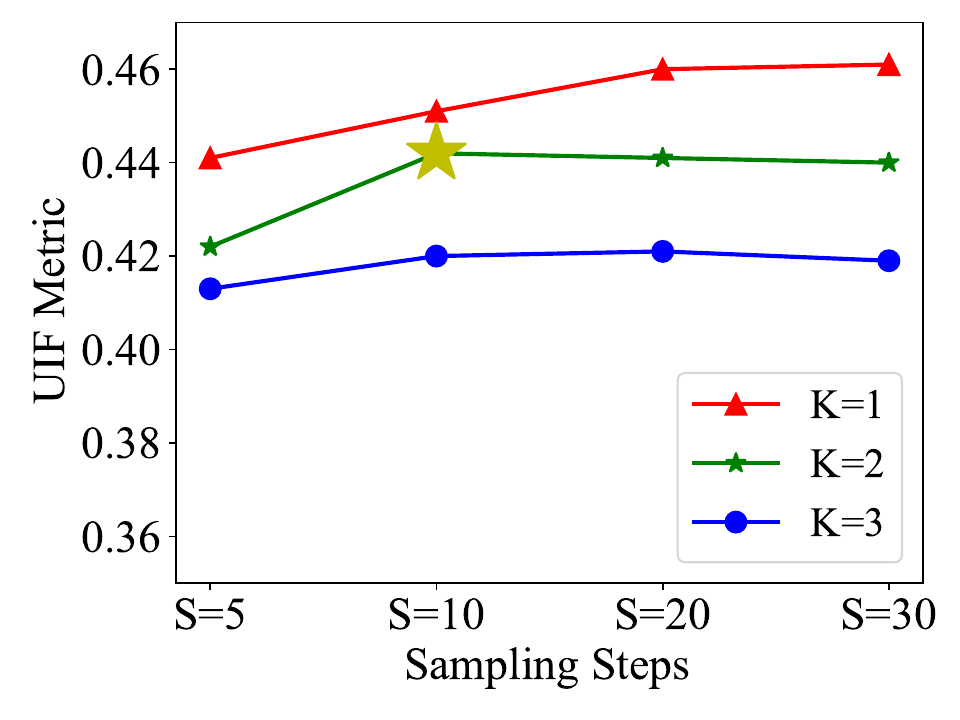}}
			\label{UIF_Metric}}
		\subfloat[Average Inference Time]{
			{\includegraphics[scale=0.276]{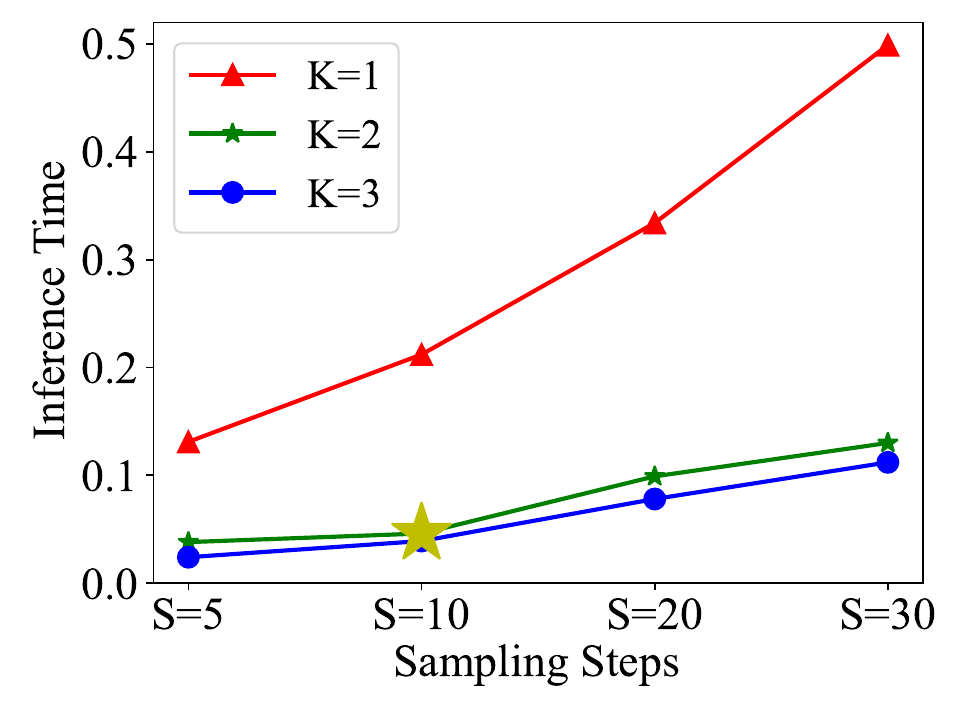}}
			\label{Time_Metric}}
		\caption{Ablation studies of wavelet transform levels $K$ and sampling steps $S$ in terms of three full-reference metrics (SSIM, PSNR, UIF) and Inference Time. The results using the default settings are marked with a yellow asterisk.}
		\label{Wavelet_Transform_Levels_and_Sampling_Step}
		\vspace{-0.2cm}
	\end{figure}
    
    \begin{figure}[!h]
		\setlength{\abovecaptionskip}{0.0cm}
		\setlength{\belowcaptionskip}{-0.2cm}
		\centering
		\includegraphics[width=0.48\textwidth]{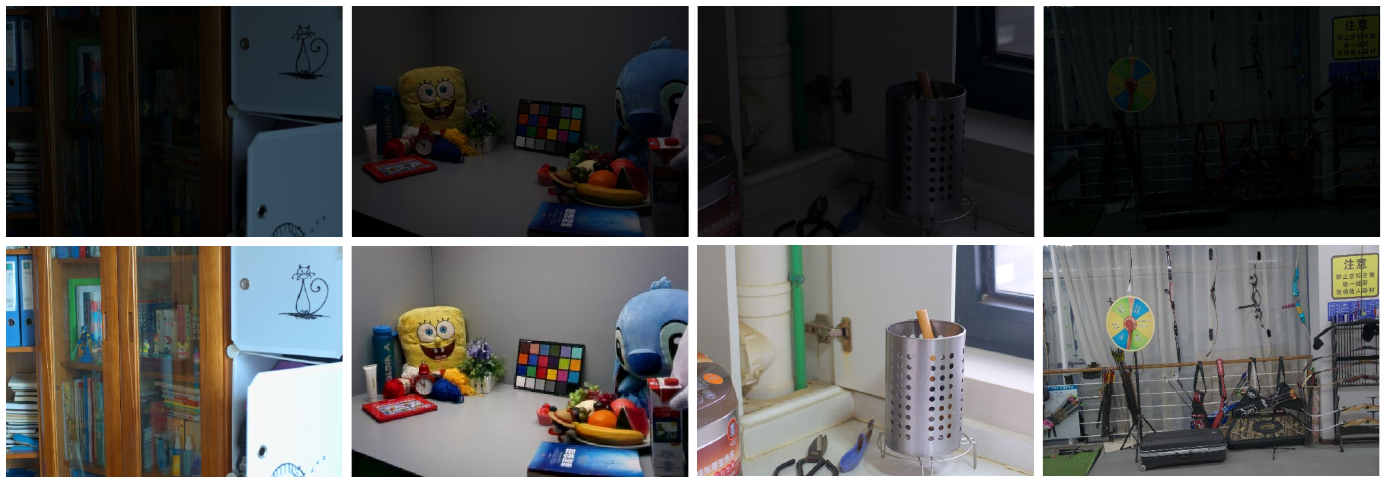}
		\caption
		{Enhanced results of low-light images from the \textbf{\textit{LOL}} dataset \cite{wei2018deep}. The top are degraded low-light images from three scenes, while the bottom are enhanced results with our DiffColor method.}
		\label{7-1Low_Light_Enhancement}
		\vspace{-0.2cm}
	\end{figure}
	
	\begin{figure}[!h]
		\setlength{\abovecaptionskip}{0.0cm}
		\setlength{\belowcaptionskip}{-0.2cm}
		\centering
		\includegraphics[width=0.48\textwidth]{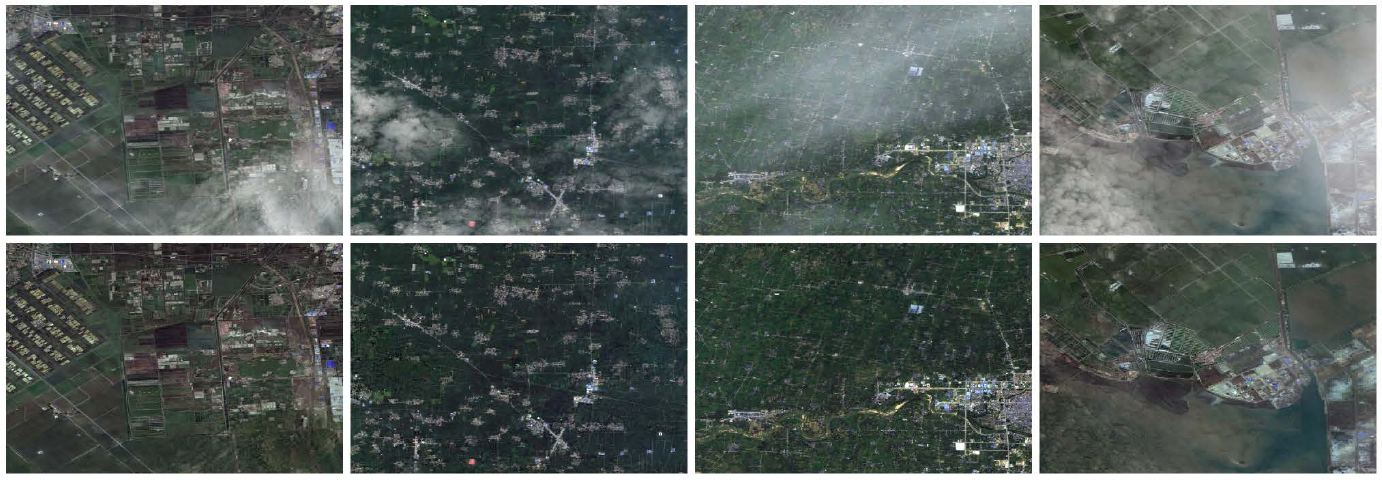}
		\caption
		{Removal results of thin cloud images from the \textbf{\textit{MCGAN}} dataset \cite{Xu2021Cloudy}. The top are four images with different cloud densities, while the bottom are cloud removal results with our DiffColor method.}
		\label{7-2Foggy_Remove}
		\vspace{-0.1cm}
	\end{figure}
    
	\textbf{High-frequency Spectrum as Conditional Input}. In Table \ref{Ablation_Studies_on_Condition}, ``-w/o HF-Conditions'' indicates that there is no refined high-frequency spectrum as input conditions for the diffusion model. Both SSIM and PSNR scores decrease compared to the full framework, demonstrating that using high-frequency signals as input conditions for the noise estimation network ensures high-fidelity content restoration even when starting from a randomly sampled noise in the reverse process.
	
	\textbf{Wavelet Transform Levels and Sampling Steps}. We further provide an ablation study about wavelet transform levels $K$ and sampling steps $S$. As shown in Fig. \ref{Wavelet_Transform_Levels_and_Sampling_Step}, we can observe that as the levels $K$ of wavelet transform increases, the reduction of image spatial resolution brings inference acceleration, but the quality of enhancement decreases in terms of SSIM, PSNR and UIF metrics. Another trend is that the increase of sampling steps $S$ leads to a clear increase in inference time, but beyond 20 sampling steps, most metrics exhibit minimal improvement. Considering the trade-off between inference time and the performance of image enhancement, we set $K=2$ and $S=10$ in the experiment.
	
	\subsection{Robustness and Generalization Validation}
	In this subsection, we also verify that the proposed DiffColor method can be directly generalized to other challenging image restoration tasks such as low-light image enhancement and thin cloud removal.
	
	Compared with low-light images, underwater images exhibit more complex and variable degradation types, such as color distortion, foggy appearance, light vertigo, low contrast, etc., making the UIE task more challenging. Fig. \ref{7-1Low_Light_Enhancement} presents the qualitative comparison results of the low-light images selected from the LOL dataset \cite{wei2018deep}. The results indicate that our DiffColor effectively improves the low-light brightness of images while ensuring color integrity. In addition, we further conducted a robustness validation of DiffColor for thin cloud removal, with the results shown in Fig. \ref{7-2Foggy_Remove}. Removing thin clouds is more challenging compared to image dehazing due to its inhomogeneous distribution property. The experimental results show that our DiffColor effectively removes obscuring clouds from the images without exhibiting excessive enhancement in cloud-free regions.
	
	\section{Conclusion}
	This paper presents DiffColor, a color distribution-aware diffusion model for underwater image enhancement. DiffColor leverages the strong generative capability of diffusion models alongside the dimensionality reduction benefits of wavelet transformation, significantly reducing the computational resource consumption of traditional conditional diffusion models. Unlike single-noise image restoration tasks, DiffColor incorporates the GCC module to address global degradation commonly encountered in reverse denoising. For the sacrificed image details caused by underwater medium scattering, DiffColor employs the CSDR module to enhance high-frequency details and utilize refined high-frequency components as the conditional signal. Through qualitative and quantitative evaluations, we validate that DiffColor achieves state-of-the-art performance in underwater image restoration.
	
	%\section*{Acknowledgments}
	%This work was supported in part by the National Natural Science Foundation of China under Grant 62225113, the Innovative Research Group Project of Hubei Province under Grant 2024AFA017, and the Australian Research Council under Project DP210101859.
	
	\bibliographystyle{IEEEtran}
	\bibliography{Ref.bib}
\end{document}